\documentclass[12pt]{iopart}
\usepackage{standalone}
\usepackage{nccfoots}
\usepackage{graphicx}
\usepackage{bbold}
\usepackage{braket}
\usepackage{color}

\definecolor{mygreen}{rgb}{0,0.5,0}
\definecolor{myblue}{rgb}{0,0,0.75}
\definecolor{mymagenta}{cmyk}{0,1,0,0.12}
\usepackage[colorlinks=true,
   linkcolor=blue,
   citecolor=blue,
   urlcolor =blue]{hyperref}

\newcommand{\citeSM}{\cite[{\tiny SM}\kern-0.3em][]{SM}}
\newcommand{\be}{\begin{equation}}
\newcommand{\ee}{\end{equation}}

\newcommand{\ie}{{\it i.e.}}

\newcommand{\beq}{\begin{eqnarray}}
\newcommand{\eeq}{\end{eqnarray}}

\def\eq#1{(\ref{#1})}
\def\H1{\widehat{H}_1}

\expandafter\let\csname equation*\endcsname\relax

\expandafter\let\csname endequation*\endcsname\relax

\usepackage{amsmath}
\usepackage{amssymb}

\begin{document}

\title[Delayed Coherent Quantum Feedback]{Delayed Coherent Quantum Feedback from a Scattering Theory and a Matrix Product State Perspective}

\author{P.-O. Guimond$^{1,2}$, M. Pletyukhov$^{3}$, H. Pichler$^{4,5}$ and P. Zoller$^{1,2}$}
\address{$^{1}$ Institute for Theoretical Physics, University of Innsbruck, A-6020, Innsbruck,
Austria \\ $^{2}$ Institute for Quantum Optics and Quantum Information of the Austrian Academy
of Sciences, A-6020 Innsbruck, Austria  \\ $^{3}$ Institute for Theory of Statistical Physics, RWTH Aachen University, 52056 Aachen, Germany\\ $^{4}$ ITAMP, Harvard-Smithsonian Center for Astrophysics, Cambridge, Massachusetts 02138, USA \\ $^{5}$ Physics Department, Harvard University, Cambridge, Massachusetts 02138, USA}

\ead{pierre-olivier.guimond@uibk.ac.at}

\begin{abstract}  
{We study the scattering of photons propagating in a semi-infinite waveguide terminated by a mirror and interacting with a quantum emitter. This paradigm constitutes an example of coherent quantum feedback, where light emitted towards the mirror gets redirected back to the emitter. We derive an analytical solution for the scattering of two-photon states, which is based on an exact resummation of the perturbative expansion of the scattering matrix, in a regime where the time delay of the coherent feedback is comparable to the timescale of the quantum emitter's dynamics. We compare the results with numerical simulations based on matrix product state techniques simulating the full dynamics of the system, and extend the study to the scattering of coherent states beyond the low-power limit.}

\end{abstract}
\date{\today}
\maketitle

\section{Introduction}
One of the paradigmatic models in quantum optics consists of a quantum emitter coupled to a one-dimensional photonic waveguide. This forms the basic block for the description of more complex quantum optical systems such as quantum networks, where multiple emitters exchange quantum information via photons propagating in one-dimensional quantum channels. 
Recent experimental realizations in the optical domain include single atoms coupled to nanofibers \cite{Mitsch:2014fz} or nanostructured waveguides \cite{Tiecke:vw,Goban:2014eq}, as well as artificial atoms such as quantum dots or color centers in diamond \cite{Sipahigil:2016hy} embedded in photonic crystal waveguides \cite{Sollner:2015fc,Lodahl:2015fy}.
The same model also describes the physics of superconducting qubits coupled to microwave transmission lines \cite{vanLoo:2013df,Hoi:2015fh}, and of systems where bosonic excitations other than photons are used, such as surface acoustic waves~\cite{PhysRevA.95.053821} or phonons in cold quantum gases \cite{Ramos:2014ut}.

From a theoretical point of view, the atom-photon interaction can often be treated in a Born-Markov approximation, and the description in terms of a master equation is excellent. However, many interesting problems require going beyond this Markovian paradigm \cite{Rivas:2014bl,Breuer:2016dn}. In particular, in the description of problems involving coherent quantum feedback the finite propagation speed of photons and the corresponding \textit{time delays} invalidate one of the assumptions of a Markovian master equation treatment \cite{QuantumNoiseBook} by introducing an effective memory. 

In recent years several theoretical methods have been developed to address different aspects of problems involving such time delays \cite{Dorner:2002dv,Grimsmo:2015gf,Pichler:2016bx,Zheng:2013fy,Laakso:2014bk,Fang:2015bc,DiazCamacho:2015tx,Ramos:2016vd,Pichler:2017wq}. 
In particular, notable progress has been made in analyzing so-called \textit{scattering problems}. There, one is interested in injecting (few) photons into a  network of distributed passive emitters and characterizing the properties of transmitted and reflected photons. While this type of problems is well studied in the Markovian regime \cite{Shen:2007wi,Shi:2009zb,Pletyukhov:2012iz,Shi:2015dc}, exact numerical and analytical solutions have been derived only recently for specific examples that include time delays \cite{Zheng:2013fy,Laakso:2014bk,Fang:2015bc}. These solutions are based on methods that are tailored to the scattering problem, such as a direct integration of the Lippmann-Schwinger equation or diagrammatic techniques. Due to the complexity increasing with the photon number, these approaches are however limited to incident states consisting of few-photon states.  

In the present work we first apply these methods to study the two-photon scattering problem. The system we consider consists of a single quantum emitter coupled to a semi-infinite 1D waveguide terminated by a mirror, and constitutes a paradigmatic example of quantum feedback in a quantum optical system, where the photons emitted towards the mirror are redirected towards the emitter. The non-Markovian element is introduced here by the finite propagation time of the coherent feedback photons relative to the timescale of the emitter's dynamics. We present analytical and numerical solutions for different types of emitters. Explicitly we consider two types of emitters: a two-level system, and a V-level system where one transition couples only to the photons propagating towards the mirror while the other transition couples only to the ones propagating outwards. This assumes chiral (\ie~unidirectional) coupling between emitter and waveguide, as has been demonstrated in recent experiments with optical setups \cite{Mitsch:2014fz,Sollner:2015fc}. Arguably, these two scenarios can be considered the most simple quantum optical examples exhibiting essential non-Markovian character due to time-delayed coherent feedback. The former because of the structure of the emitter, the latter because an incoming photon scattering on the emitter will necessarily undergo only a single roundtrip between the emitter and the mirror before leaving the system \cite{Guimond:2016kf}. 

In a different approach, some of us have recently developed numerical techniques to describe the \textit{full time-dependent dynamics} of photonic quantum networks with time delays \cite{Pichler:2016bx}. This method makes use of matrix product state techniques \cite{Vidal:2003gb,Daley:2004hk,Verstraete:2010bf,Schollwock:2011gl} to represent the entangled state of the propagating photons. Due to this efficient representation of the effective many-body Hilbert space these methods are not limited by the number of photons, and can be used to analyze systems where the photon number is not conserved. This includes the important case  of coherent input fields, with a finite photon intensity. In this work we make use of these techniques to study the scattering of coherent states, beyond the low-power regime usually assumed when using analytical methods. Further we verify that simulations of two-photon state scattering yield results in agreement with the analytical predictions.

The paper is organized as follows. In section~\ref{sec:scattering} we formally introduce our model and formulate the scattering problem. In section \ref{sec:diagrammatic} we develop our analytical approach for the scattering of two-photon states, performing a resummation of the expansion of the scattering matrix. Section~\ref{sec:mps} provides details on the numerical approach. In section \ref{sec:Results} we discuss the solutions obtained with both techniques. We present the solution of the two-photon scattering problem, and extend the study to coherent driving fields, discussing the limit in which an analytical approximation based on a truncation of the driving field to few-photon components captures the results.

\begin{figure}
\begin{center}
\includegraphics[width=14cm]{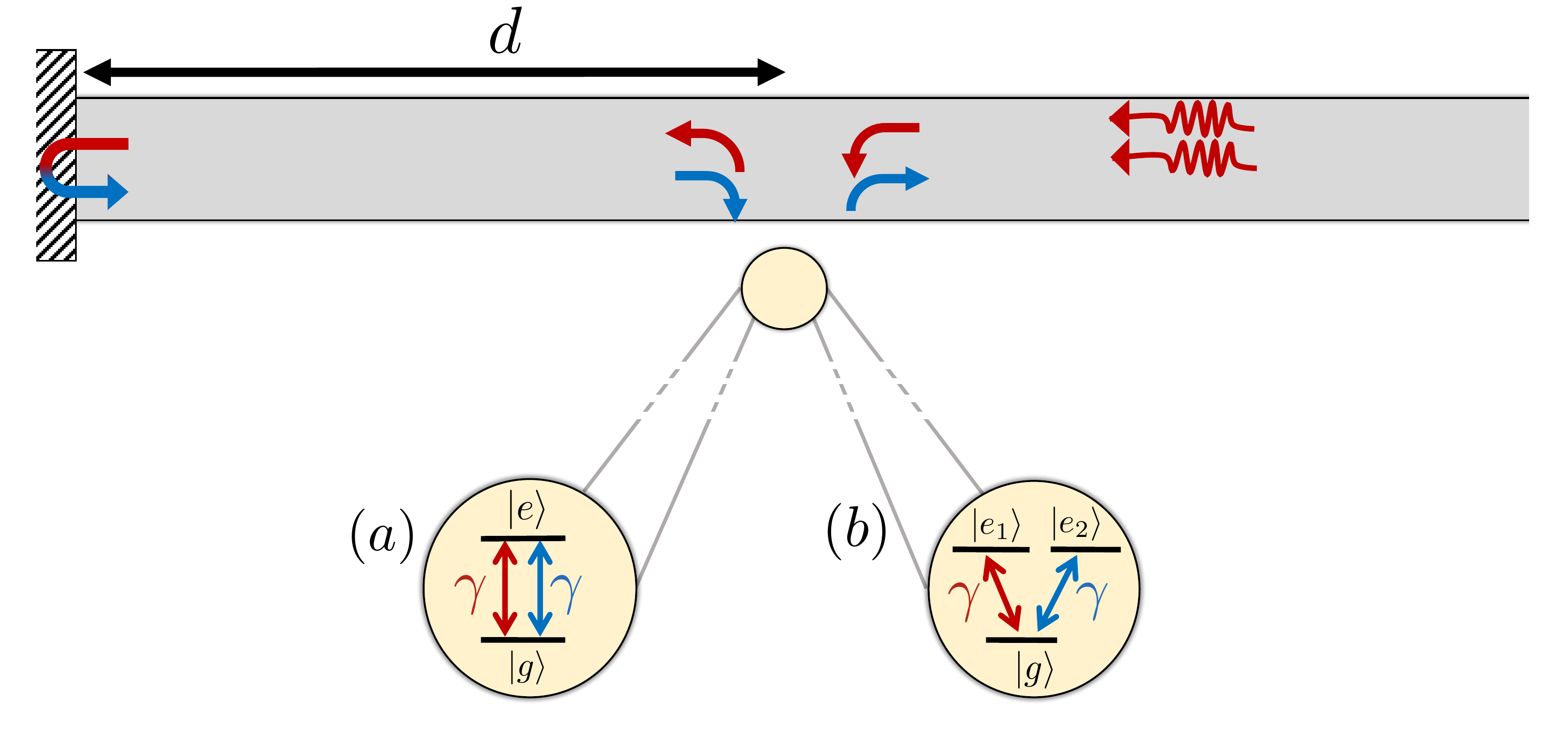}
\caption{\label{fig1} Schematic of the setup. Incoming photons scatter off of a quantum emitter coupled to a one-dimensional waveguide and located at a distance $d$ from a mirror. We consider the following two examples. (a) The atom realizes a two-level system. (b)~Chiral setup where the atom realizes a V-level system with one transition coupling to the waveguide modes propagating to the left and one transition coupling to the ones propagating to the right.
 }
\end{center}
\end{figure}

\section{Model}

\label{sec:scattering}
\subsection{Hamiltonian}

The system we consider is depicted in Fig.~\ref{fig1}, where an atom as quantum emitter is coupled to a semi-infinite waveguide terminated by a mirror. The dynamics of the system of atom and waveguide is dictated by the Hamiltonian $H=H_a + H_B+V$ \cite{Dorner:2002dv}, where $H_a$ is the free atomic Hamiltonian (discussed below), $H_B=\int d\omega\,\omega\, b^\dagger(\omega)b(\omega)$ is the free waveguide Hamiltonian (in units with $\hbar=1$), and 
\begin{align}\label{eq:InteractionHamiltonian}
  V= i \sqrt{\gamma/2\pi} \int d\omega \big(b^\dagger(\omega)(\sigma_1 e^{i\omega \tau/2}-\sigma_2 e^{-i\omega \tau/2}) -\text{H.C.}  \big),\end{align} describes the coupling of the atom to the waveguide. Here the integrals run over a broad bandwidth around a characteristic optical frequency $\bar\omega$ close to the frequencies of the scattered photons and the atomic transitions. The operator $b^\dagger(\omega)$ is the creation operator for a photon with frequency $\omega$ in the waveguide, satisfying \mbox{$[b(\omega),b^\dagger(\omega')]=\delta(\omega-\omega')$}, and $\sigma_1$ and $\sigma_2$ are the atomic transition operators coupling to left and right moving photons respectively \cite{Pichler:2016bx}. The finite time delay of the coherent feedback \mbox{$\tau=2d/c>0$} introduces frequency-dependent phase factors in the coupling strength. The interaction Hamiltonian \eqref{eq:InteractionHamiltonian} assumes a linear dispersion relation $\omega(k)\approx c|k|$ over the relevant bandwidth, with the speed of light in the waveguide denoted by $c$  and the distance of the atom to the mirror by $d$. The opposite sign in the coupling to left- and right-moving photons accounts for the $\pi$-shift in the phase of the photons reflected by the mirror. We also made use of the rotating wave approximation (RWA) and considered the absolute values of the coupling strengths, $\gamma$, to be independent of the frequency in the relevant bandwidth. Moreover, we neglect waveguide losses and couplings of the atom to modes outside of the waveguide.


We consider two examples for the atomic structure and its coupling to the waveguide. First, we consider an atom that is described by a two-level system (TLS) with ground state $\ket{g}$ and excited state $\ket{e}$ (see Fig.~~\ref{fig1}(a)). We naturally assume that this two-level system couples equally to both left and right moving photons. In this case, the atomic operators are given by $\sigma_1 = \sigma_2 = \ket{g}\bra{e}$ and $H_a=\omega_a \ket{e}\bra{e}$, where $\omega_a$ is the two-level transition frequency. Since the excited state $\ket{e}$ can decay by emission of photons into both directions, its total decay rate is $2\gamma$  (in the absence of the mirror). In the second example, the atomic level structure is represented by a V-level system with ground state $\ket{g}$ and excited states $\ket{e_1}$ and $\ket{e_2}$ (see Fig.~~\ref{fig1}(b)). In this case the atomic operators are defined as $\sigma_1=\ket{g}\bra{e_1}, \sigma_2=\ket{g}\bra{e_2}$ and $H_a=\omega_1 \ket{e_1}\bra{e_1} + \omega_2 \ket{e_2}\bra{e_2}$, where $\omega_{1,2}$ is the frequency of each atomic transition. Note that in this example photons emitted from the atom in state $\ket{e}_1$ propagate towards the mirror, while photons emitted from state $\ket{e_2}$ propagate away from the mirror, as realized by a chiral coupling \cite{Lodahl:2017bz}. Thus, in this case both excited states decay with a rate $\gamma$ (in the absence of the mirror).
In the following we will use a generic notation for these two examples, denoting the excited states as $\ket{e_\chi}$, with $\chi$ labelling the transitions, and writing the Hamiltonian as $H_a=\sum_\chi \omega_\chi \ket{e_\chi}\bra{e_\chi}$.

For convenience we move to a frame rotating with the characteristic frequency $\bar\omega$. In addition we rescale the photon frequencies as $\omega-\bar\omega\to\nu$, and correspondingly redefine $b(\omega)\to b(\nu)$. In this frame the free Hamiltonians for the atom and the waveguide now read 
\be \label{eq:defHaB}H_a=-\sum_{\chi} \delta_\chi \ket{e_\chi}\bra{e_\chi} \text{ and } H_B = \int d\nu\, \nu\, b^\dagger(\nu)b(\nu), \ee 
where the detuning is $\delta_\chi=\bar\omega-\omega_\chi$ and the interaction Hamiltonian reads 
\be\label{eq:defVt} V= i \sqrt{\gamma/2\pi} \int d\nu \big(b^\dagger(\nu)(\sigma_1 e^{i(\bar\omega+\nu) \tau/2}-\sigma_2 e^{-i(\bar\omega+\nu) \tau/2}) -\text{H.C.}  \big).\ee
We can identify two physically relevant dimensionless quantities characterizing the delayed interactions in this system. These are the delay-bandwidth product $\gamma\tau$, quantifying the degree of non-Markovianity, and the roundtrip propagation phase that a photon acquires  when traveling from the atom to the mirror and back, $\phi\equiv\bar\omega\tau+\pi$ (modulo $2\pi$).

\subsection{Scattering problem}

In the scattering problem we consider a state $\ket{\Psi_i}$ of free incident photons propagating in the waveguide towards the quantum emitter that is initially in the ground state. After the interaction with the emitter, the photons are scattered back into the waveguide in a final state $\ket{\Psi_f}$ with the emitter returning to its initial state. The goal is then to calculate this output state, \ie~the \emph{scattering operator} $S$, with $\ket{\Psi_f}=S\ket{\Psi_i}$. Its matrix elements are typically computed in the eigenbasis of $H_B$, which is the basis of states with definite photon numbers and frequencies and corresponding to infinitely extended wave packets. In practice we approximate an incident photon of a given frequency $\nu_1$ by a wave packet with a finite length $L$, e.g. by the square pulse with spectrum \be\label{eq:defphi}\phi_{\nu_1}(\nu)=\sqrt{\frac{2c}{\pi L}}\frac{\sin((\nu-\nu_1) L/2c)}{\nu-\nu_1}, \ee which approaches $\sqrt{\frac{2 \pi c}{L}} \delta(\nu-\nu_1)$ for $L$ large. For the scattering of incoming states consisting of two such photons with frequency $\nu_1$, the initial state is of the form \mbox{$\ket{\Psi_i}=\frac{1}{\sqrt{2}}\big(\int d\nu \phi_{\nu_1}(\nu)b^\dagger(\nu)\big)^2\ket{g,\text{vac}}$}, where $\ket{g,\text{vac}}$ corresponds to the ground state for the emitter and the vacuum state for the waveguide. Due to the fact that the number of excitations is conserved by the Hamiltonian in the RWA, the final state then also consists of two photons in the waveguide.

\section{Scattering theory approach}
\label{sec:diagrammatic}
\subsection{Scattering matrix}

In the interaction picture with $H_0\equiv H_a+H_B$, the unitary evolution from time $t'$ to $t$ is given by the operator \be U(t,t') = \mathcal T \text{exp}\Big(-i\int_{t'}^t dt'' V(t'') \Big),\ee where $\mathcal T$ denotes the time-ordering superoperator, and $V(t)=e^{iH_0t}Ve^{-iH_0 t}$. The scattering operator is then defined as $S\equiv\lim_{t\to\infty} U(t,-t)$.
This scattering operator is conveniently expressed as \cite{ScatteringBook}
\be \label{eq:defscatmat}S=\mathbb{1} -2\pi i\, \delta(E_f-E_i)\, T(E_i).\ee  
Here $E_i$ and $E_f$ are the energies of the initial and final state, \ie~$H_0\ket{\psi_i}=E_i\ket{\psi_i}$ and $H_0\ket{\psi_f}=E_f\ket{\psi_f}$. The $\delta-$function ensures energy conservation in the scattering process. The first term in \eq{eq:defscatmat} describes the free evolution, while the interacting part is contained in the \emph{T-operator} $T(E)$, which is defined with the infinite perturbative series
\be\label{eq:pertTE} T(E)=\sum_{n=0}^\infty V\left(\frac{1}{E-H_0 + i\eta} V\right)^n\ee
with $\eta=0^+$ and $V$ given in \eqref{eq:defVt}.
Since we consider problems with a conserved number of excitations, we can decompose the scattering operator into a direct sum $S =\bigoplus_{N=0}^{\infty} S^{(N)}$, where each term $S^{(N)}$ acts solely in the corresponding $N$-photon subspace. Analogously we can decompose $T(E)=\bigoplus_{N=0}^{\infty} T^{(N)}(E)$. 
Below we explicitly calculate analytical expressions for matrix elements of $T^{(1)}(E)$ and $T^{(2)}(E)$.  
To keep notation in the following derivation as compact as possible we abbreviate the interaction Hamiltonian by $V=b^\dagger_\nu v_\nu+v_\nu^\dagger b_\nu$, where $b_\nu\equiv b(\nu)$, and 
\mbox{$v_\nu=i\sqrt{\gamma/2\pi} (\sigma_1e^{i(\bar\omega+\nu)\tau/2}-\sigma_2e^{-i(\bar\omega+\nu)\tau/2})$}
is the bare interaction vertex. The integration over repeated $\nu$ is implicit. 

\subsubsection{Single-photon subspace. ---}
We first start with the simpler case of a single-photon scattering ($N=1$). Using the expansion \eqref{eq:pertTE}, we can calculate all matrix elements in this sector as 
\begin{align}  \bra{g,\nu'_1}T^{(1)}(E)\ket{g,\nu_1}&=\bra{g,\nu_1'} \sum_{n=0}^\infty V\left(\frac{1}{E-H_0 + i\eta} V\right)^n \ket{g,\nu_1} \\  &= \nonumber\bra{g,\nu_1'} V \sum_{n=0}^\infty \left(\frac{1}{E-H_0 + i\eta} V\right)^n\frac{1}{E-H_0 + i\eta} V \ket{g,\nu_1} \\ &=\nonumber \sum_{\chi',\chi} g^*_{\chi'}(\nu_1') \bra{e_{\chi'},\text{vac}}\sum_{n=0}^\infty \left(\frac{1}{E-H_0 + i\eta} V\right)^n\ket{e_\chi,\text{vac}}\frac{g_\chi(\nu_1)}{E+\delta_\chi + i\eta}  
\\ \nonumber&\equiv\sum_{\chi',\chi} {g_\chi(\nu_1)g^*_{\chi'}(\nu_1')}  \bra{e_{\chi'}}M(E)\ket{e_\chi}
,\end{align}
where $\ket{g,\nu}$ denotes the state with the atom in the ground state and a single photon with frequency $\nu$ in the waveguide. In the second line we used $\bra{g,\nu'}V\ket{g,\nu}=0$ and shifted the sum. In the third line we defined \be \label{eq:defgbetanu} g_\chi(\nu)\equiv \bra{e_\chi}v^\dagger_\nu\ket{g}=-i\sqrt{\gamma/2\pi}\left(\bra{e_{\chi}}\sigma_1^\dagger\ket{g} e^{-i(\bar\omega +\nu)\tau/2}-\bra{e_{\chi}}\sigma_2^\dagger\ket{g} e^{i(\bar\omega +\nu)\tau/2}\right),\ee and in the fourth line we defined the dressed atomic Green's function projected on the excited subspace 
\be M(E)= \sum_{n=0}^\infty \bra{\text{vac}}\left(\frac{1}{E-H_0 + i\eta} V\right)^n\ket{\text{vac}} M^0(E),\ee
with $M^0(E)=\sum_\chi \frac{1}{E+\delta_\chi + i\eta} \ket{e_\chi}\bra{e_\chi}$ the corresponding bare atomic Green's function.
To obtain an explicit expression for $M(E)$, we note that only the even powers of $n$ contribute to the sum, since $\bra{\text{vac}}V\ket{\text{vac}}=0$. We then obtain the Dyson equation
\begin{align} \label{eq:dyson}M(E) &= \sum_{n=0}^\infty \bra{\text{vac}}\left(\frac{1}{E-H_0 + i\eta} V\frac{1}{E-H_0 + i\eta} V\right)^n\ket{\text{vac}} M^0(E)
\\\nonumber &= \sum_{n=0}^\infty \left(M^0(E) \bra{\text{vac}}V\frac{1}{E-H_0 + i\eta} V\ket{\text{vac}}\right)^n M^0(E)\\
\nonumber &= \sum_{n=0}^\infty \left(M^0(E) \Sigma(E)\right)^n M^0(E)=\frac{1}{1-M^0(E)\Sigma(E)}M^0(E).
\end{align}
In the last line we defined the \emph{self-energy} of the atom
\begin{align}\label{eq:defSigma} \Sigma(E)&=\bra{\text{vac}}V\frac{1}{E-H_0+i\eta}V\ket{\text{vac}}
= \sum_{\chi,\chi'}\int d\nu \frac{g_\chi(\nu) g_{\chi'}^*(\nu)}{E-\nu+i\eta} \ket{e_\chi}\bra{e_{\chi'}}
\\ \nonumber&= -{i}\gamma\sum_{\chi,\chi'}\left(\frac12\bra{e_\chi}(\sigma_1^\dagger\sigma_1+\sigma_2^\dagger\sigma_2)\ket{e_\chi'} - e^{i(\bar\omega + E)\tau}\bra{e_\chi}\sigma_2^\dagger\sigma_1\ket{e_{\chi'}}\right) \ket{e_{\chi}}\bra{e_{\chi'}}, \end{align}
where we used the explicit expression for $g_\chi(\nu)$ given in \eqref{eq:defgbetanu}. The self-energy is interpreted as the energy of the atom arising from the interaction with the waveguide, restricted to the excited subspace. The first component is purely imaginary and corresponds here to the deterministic decay of an excitation via photon emission at the rate $\gamma$. The second component on the other hand is the dipole-dipole interaction between the two transitions, which depicts the driving of the second transition by the first one via the delayed exchange of photons with frequency $\bar\omega+E$. Note that in the Markovian limit, that is, for $\gamma\ll1/\tau$, the phase factor $e^{i(\bar\omega+E)\tau}$ in \eqref{eq:defSigma} is  approximately constant over the relevant bandwidth, and $\Sigma(E)$ becomes independent of the frequency $E$.
Using \eqref{eq:dyson} and \eqref{eq:defSigma} we finally obtain the single-photon scattering matrix from \eqref{eq:defscatmat} via
\begin{align}
\langle g |S^{(1)}_{\nu'_1 , \nu_1} | g\rangle = \delta_{\nu'_1 \nu_1} \left( 1 -2 \pi i \langle g | v_{\nu_1} M (\nu_1) v_{\nu_1}^{\dagger} | g\rangle \right).
\end{align}

\subsubsection{Two-photon subspace. ---}
We now apply the same approach for two-photon scattered states ($N=2$). Let us consider an input state $\ket{\Psi_i}$ and an output state $\ket{\Psi_f}$ with the atom in the ground state and two photons in the waveguide. The expansion \eqref{eq:pertTE} then reads
\begin{align} \bra{\Psi_f}T^{(2)}(E)\ket{\Psi_i}&=\bra{\Psi_f} \sum_{n=0}^\infty V\left(\frac{1}{E-H_0 + i\eta} V\right)^n \ket{\Psi_i} 
\\  &= \nonumber\bra{\Psi_f} V \sum_{n=0}^\infty \left(\frac{1}{E-H_0 + i\eta} V\right)^n\frac{1}{E-H_0 + i\eta} V \ket{\Psi_i}
.\end{align}
Using the fact that $V\ket{\Psi_i}$ has one photon and belongs to the atomic excited subspace, and noting that once again only even values of $n$ contribute to the sum, this expression can be rewritten as
\begin{align} \label{eq:reecT}
\bra{\Psi_f} V \sum_{n=0}^\infty \left(M^0(E-H_B) V\frac{1}{E-H_0 + i\eta} V\right)^nM^0(E-H_B) V \ket{\Psi_i},
\end{align}
where $M^0$ is now evaluated at $E-H_B$ due to the presence of an additional photon. Let us now consider the term \mbox{$V\frac{1}{E-H_0+i\eta}V$}, which in \eqref{eq:reecT} is also an operator acting on the subspace with one photon and the atom in an excited state.
Its matrix elements are given by
\begin{align} \bra{e_{\chi'},\nu'}&V\frac{1}{E-H_0+i\eta}V\ket{e_\chi,\nu}
=\bra{e_{\chi'},\text{vac}}b_{\nu'}v_{\nu_3}^\dagger b_{\nu_3} \frac{1}{E-H_0+i\eta}b_{\nu_4}^\dagger v_{\nu_4} b^\dagger_\nu\ket{e_\chi,\text{vac}}
\\\nonumber&=\bra{e_{\chi'}}v_{\nu_3}^\dagger v_{\nu_4} \ket{e_\chi}\frac{\delta_{\nu_3,\nu_4}\delta_{\nu,\nu'}+\delta_{\nu',\nu_4}\delta_{\nu,\nu_3}}{E-\nu_4-\nu+i\eta}
= \bra{e_{\chi'},\nu'}\Sigma(E-H_B) + \Omega(E)\ket{e_\chi,\nu}.  \end{align} 
Here $\Sigma(E)$ corresponds to the path without interaction between the atom and the photon, where $\nu'=\nu$ and $\nu_3=\nu_4$, and is defined in \eqref{eq:defSigma}. On the other hand, $\Omega(E)$ corresponds to the path with interaction, with $\nu_4=\nu'$ and $\nu_3=\nu$, and is expressed as 
\be \label{eq:defOmega}\Omega(E)=b^\dagger_{\nu'} v^\dagger_\nu \frac{1}{E-\nu-\nu'+i\eta}v_{\nu'} b_\nu. \ee
The $T$-operator in \eqref{eq:reecT} can then be written as
\begin{align}\nonumber
\bra{\Psi_f}T^{(2)}(E)\ket{\Psi_i} =&\bra{\Psi_f}V\frac{1}{1-M^0(E-H_B)(\Sigma(E-H_B)+\Omega(E))} M^0(E-H_B)V\ket{\Psi_i} 
\\ \nonumber =&\bra{\Psi_f}V\frac{1}{1-M(E-H_B)\Omega(E)} M(E-H_B)V\ket{\Psi_i} 
 \\ \nonumber =&\bra{\Psi_f}V\sum_{n=0}^\infty\Big(M(E-H_B)\Omega(E)\Big)^n M(E-H_B)V\ket{\Psi_i} 
 \\ \nonumber =&\bra{\Psi_f}VM(E-H_B)V\ket{\Psi_i} \\ &+\bra{\Psi_f} VM(E-H_B)W(E)M(E-H_B)V\ket{\Psi_i} ,
 \end{align}
where in the second line we made use of the Dyson equation \eqref{eq:dyson}, and $W(E)$ is defined implicitly via  the recursive relation 
\be\label{eq:recWE} W(E)=\Omega(E)+\Omega(E)M(E-H_B) W(E). \ee 


Finally, from \eqref{eq:defscatmat} we obtain that the coefficients of the two-photon scattering matrix are given by 
\begin{align}
\langle g |S^{(2)}_{\nu'_1 \nu'_2, \nu_1 \nu_2}| g \rangle =&\frac12  \delta_{\nu'_1 \nu_1}  \delta_{\nu'_2 \nu_2} - 2 \pi i \bra{g}v_{\nu'_1} M (\nu_1) v_{\nu_1}^{\dagger}\ket{g} \delta_{\nu'_1 \nu_1}
\delta_{\nu'_2 \nu_2} \nonumber \\
&- 2 \pi i \bra{g}v_{\nu'_1} M (\nu'_1) W_{\nu'_2,\nu_2}(\nu_1+\nu_2) M (\nu_1) v_{\nu_1}^{\dagger}\ket{g} \delta_{\nu'_1+\nu'_2 , \nu_1 + \nu_2},
\label{S2}
\end{align}
where $W_{\nu',\nu}(E)\equiv \bra{\nu'}W(E)\ket{\nu}$. From \eqref{eq:defOmega}, the contribution $\bra{\nu'}\Omega(E)\ket{\nu}$ to \eqref{eq:recWE} can be decomposed into the elastic part $-i \pi v_{\nu}^{\dagger} v_{\nu'} \delta (E- \nu - \nu') $ and the inelastic part $\mathcal{P}v_{\nu}^{\dagger} \frac{1}{E -\nu - \nu'} v_{\nu'}$, with $\mathcal P$ the principal value. We can then alternatively represent \eqref{S2} by
\begin{align}
\langle g |S^{(2)}_{\nu'_1 \nu'_2, \nu_1 \nu_2}| g \rangle &= \frac12  \langle g |S^{(1)}_{\nu'_1,\nu_1}| g \rangle  \langle g |S^{(1)}_{\nu'_2, \nu_2}| g \rangle  
\nonumber \\
&- 2 \pi i \bra{g}v_{\nu'_1} M (\nu'_1) W_{\nu'_2, \nu_2}^{\mathcal{P}} (\nu_1+\nu_2) M (\nu_1) v_{\nu_1}^{\dagger}\ket{g} \delta_{\nu'_1+\nu'_2 , \nu_1 + \nu_2},
\label{S2alt}
\end{align}
where $ W_{\nu',\nu}^{\mathcal{P}} (E) \equiv  W_{\nu', \nu} (E) +i \pi  v_{\nu}^{\dagger} v_{\nu'} \delta (E- \nu - \nu')$. This representation is convenient for the clear separation between elastic and inelastic scattering processes, respectively in the first and second term of \eqref{S2alt}.

\subsection{Observables}

In the following we will consider an initial state consisting of two photons with the same frequency, to which we will set the rotating frame frequency $\bar\omega$. The initial state is approximated by the wavepacket $\ket{\Psi_i}=\frac{1}{\sqrt{2}}\big( \int d\nu\, \phi_0(\nu) b^\dagger(\nu)\big)^2 \ket{g,\text{vac}},$
with $\phi_0(\nu)$ defined in \eqref{eq:defphi}. Using the relation given in eq.~\eqref{S2alt}, the final state can be expressed as
\begin{align}\label{eq:s2psiobs}
\ket{\Psi_f}= S^{(2)}\ket{\Psi_i} = \frac1{\sqrt{2}} \Big(\int d\nu\, s\, \phi_0(\nu)b^\dagger(\nu)\Big)^2 |g,\text{vac} \rangle 
+ \frac{c}{\sqrt{2} L} \int d \nu \,\hat{t}(\nu)  b^\dagger(\nu) b^\dagger(-\nu) |g,\text{vac} \rangle,
\end{align}
where we defined 
\be \label{eq:stnudef}
\begin{aligned}
s &= 1 - 2 \pi i \langle g | v_0 M(0) v_0^{\dagger} | g \rangle,\quad \textrm{and}\quad \hat{t}(\nu) = \frac12 \left[ t(\nu) + t(-\nu) \right],
\end{aligned}
\ee
with $t(\nu) = - 8 i \pi^2 \langle g | v_{\nu} M (\nu) W^{\mathcal{P}}_{- \nu,0} (0) M(0) v_{0}^{\dagger} | g \rangle$.
Here $s$ is a phase factor acquired by each photon during an elastic scattering, and the second term contains the inelastic part which generates entangled states of photons with opposite frequencies. In the derivation of the second term of \eqref{eq:s2psiobs} we have assumed the limit $L \to \infty$, hence replaced $\phi_0 (\nu) \to \sqrt{\frac{2 \pi c}{L}} \delta (\nu)$. However, we have retained the full form of $\phi$ in the first term, as it will be needed to regularize  the expression of observables in which products of the form $\phi_0^* (\nu) \phi_0 (\nu)$ appear. In this situation it is necessary to employ the normalization condition $\int d \nu |\phi_0 (\nu) |^2 =1$. Below we express observables as a function of $s$ and $\hat t(\nu)$. 

\subsubsection{Power spectrum. ---}

From \eqref{eq:s2psiobs}, we find 
\begin{align}
b(x)  | \Psi_{f} \rangle =&  \sqrt{\frac{2}{L}}s^2\int d\nu\, \phi_0(\nu) b^\dagger (\nu) |g,\text{vac} \rangle + \frac{\sqrt{c}}{L \sqrt{\pi}} \int d\nu\, e^{i \nu x/c }\, \hat{t}(\nu) b^\dagger(-\nu)  |g,\text{vac} \rangle, 
\label{aPsi}
\end{align}
where \mbox{$
b(x) \equiv \frac{1}{\sqrt{2 \pi c}}  \int d\nu\, b_{\nu} e^{i\nu x/c}
$} is the field operator in the position representation.
We then obtain the first-order field correlation as
\begin{equation} \label{eq:autoG1}G^{(1)}(t)\equiv\, c\bra{\Psi_f}b^\dagger(x-ct)b(x)\ket{\Psi_f} = \frac{2 c }{L}+ \frac{4c^2}{L^2}\text{Re}[(s^*)^2 \hat t(0)] + \frac{c^2}{L^2 \pi}\int d\nu\,e^{i\nu t} |\hat t(\nu)|^2. \end{equation}
The Fourier transform of this expression yields the elastic and inelastic power spectra, which read
\begin{equation}
\label{Sel_inel}S_\text{el}(\nu)= \Bigg(\frac{2 c }{L}+ \frac{4c^2}{L^2}\text{Re}[(s^*)^2 \hat t(0)] \Bigg) \delta(\nu),
\ S_\text{inel}(\nu)= \frac{c^2}{L^2\pi}|\hat t(\nu)|^2, 
\end{equation}
where the elastic part is obtained from the first two terms of \eqref{eq:autoG1}, which are time-independent, and the inelastic part from the last term.

\subsubsection{Second order field correlation. ---}

Similarly, we establish
\be
b(x-ct)b(x)\ket{\Psi_f}=\frac{\sqrt{2}}{L}\Big[ s^2 + \frac1{2\pi}\int d\nu\, e^{i\nu t}\hat t(\nu) \Big]\ket{\text{vac}}. 
\ee
Combining this result with \eqref{aPsi}, we obtain the normalized second order auto-correlation function, which reads (in the limit $L\to\infty$)
\begin{equation}
g^{(2)} (t) = \frac{\langle  \Psi_{f}  | b^{\dagger} (x)  b^{\dagger}(x- c t) b(x- c t) b(x) |  \Psi_{f}  \rangle }{\langle   \Psi_{f}  | b^{\dagger} (x)  b (x) |  \Psi_{f}   \rangle ^2} = \frac12\bigg| 1 +  \frac{1}{2 \pi s^2} \int d \nu  \, \hat{t}(\nu) \cos(\nu t) \bigg|^2.
\label{g2def}
\end{equation}

{We have thus derived analytical expressions for the observables in our scattering problem in terms of the components of the scattering matrix. The explicit expressions for the cases where the quantum emitter is a two-level or a V-level system are given in \ref{app:explicit}.
}

\subsection{Scattering of coherent states in the low-power limit}
In quantum optical experiments one typically studies the scattering of coherent states. There the photon number is not conserved, and the input states contain a constant photon intensity rather than a fixed photon number. The previous results can be easily adapted to the scattering of coherent states with low amplitude, as only the few-photon components contribute to the dynamics in the limit of weak coherent driving fields. In this regime we can expand the weak coherent input state in different photon number sectors. Keeping only contributions up to two-photon states we can approximate a weak coherent field with amplitude $\alpha$ as
\begin{align}
	\ket{\Psi_i}=\, &e^{-|\alpha|^2/2}\sum_{n=0}^\infty\frac{\alpha^n}{n!}\Big(\int d\nu\,\phi_0(\nu) b^\dagger(\nu)\Big)^n \ket{g,\text{vac}} \\\approx\, &\nonumber e^{-|\alpha|^2/2}\Big[1 + \alpha \int d\nu\,\phi_0(\nu) b^\dagger(\nu)+ \frac{\alpha^2}{2}\Big(\int d\nu\,\phi_0(\nu) b^\dagger(\nu)\Big)^2\Big]\ket{g,\text{vac}}.
\end{align}
With the scattering theory approach we can calculate the corresponding output state in terms of the variables \eqref{eq:stnudef},
\begin{align}
	\ket{\Psi_f}\approx  e^{-|\alpha|^2/2}\Big[1 &+ \alpha s \int d\nu\,\phi_0(\nu) b^\dagger(\nu)\\&+ \nonumber\frac{\alpha^2}{2}\Big(\int d\nu\,s\,\phi_0(\nu) b^\dagger(\nu)\Big)^2 + \frac{c}{2L}\int d\nu\, \hat t(\nu)b^\dagger({\nu})b^\dagger({-\nu})\Big]\ket{g,\text{vac}}.
\end{align}
The previous analytical derivations can then be readily reproduced using these states. In terms of \eqref{Sel_inel}, the inelastic part of the power spectrum then reads $S_\text{inel}^\text{coh}(\nu)=(|\alpha|^4/2) S_\text{inel},$ to leading order in $|\alpha|$. Conversely, the second order field correlation from \eqref{g2def} now reads $g^{(2)}_\text{coh}(t)=2 g^{(2)}(t)$.

\section{Matrix product state approach}
\label{sec:mps}
We now turn to our numerical approach, which employs matrix product state (MPS) techniques to simulate the evolution of the entangled state of photonic field and emitter, allowing to go beyond the two photon limit. It is based on the methods developed in \cite{Pichler:2016bx} where the authors studied non-Markovian dynamics of quantum systems strongly driven by a continuous classical field. Here we adapt these methods in order to calculate also the scattering of states with fixed photon number. 

\subsection{Stroboscopic evolution}
In order to apply our matrix product state approach we cast the Hamiltonian given in \eqref{eq:defHaB} and \eqref{eq:defVt} in the interaction picture with respect to $H_B$. Moreover, we define quantum noise operators as 
\mbox{$b(t)=\frac{1}{\sqrt{2\pi}}\int d\nu\, b(\nu) e^{-i\nu t}$}
such that the Hamiltonian now depends explicitly on time and becomes 
\be H(t)=H_a+i \sqrt{\gamma}\big( b^\dagger(t+\tau/2) \sigma_1e^{i\bar\omega \tau/2}-b^\dagger(t-\tau/2) \sigma_2 e^{-i\bar\omega \tau/2} -\text{H.C.}  \big), \ee 
or, under the transformation \mbox{$b(t)\to b(t-\tau/2)e^{i\bar\omega\tau/2}$},
\be H(t)=H_a+i \sqrt{\gamma}\big( b^\dagger(t) \sigma_1-b^\dagger(t-\tau) \sigma_2 e^{-i\bar\omega \tau} -\text{H.C.}  \big). \ee 
The dynamics of the system is then given by the Quantum Stochastic Schr\"odinger Equation (QSSE) $i \frac{d}{dt} \ket{\Psi(t)} = H(t) \ket{\Psi(t)}.$

The MPS-algorithm approximates the dynamics as a stroboscopic evolution at discrete times $t_{k}=t_{k-1}+\Delta t$ for $k=1,..., N$ with $t_0$ the initial time, $\Delta t$ the time step interval and $N$ the total number of time steps during the evolution. In order to approximate the continuous evolution we choose a step size that is much smaller than the time scales on which the atom evolves, \ie~$\gamma\Delta t\ll1$. In each time-step $k$ we define operators $\Delta B_k = \int_{t_{k-1}}^{t_{k}}dt\, b(t),$ which are known as {\it noise increments} in the context of quantum stochastic calculus and satisfy bosonic commutation relations $[\Delta B_k, \Delta B_{k'}^\dagger]=\Delta t\, \delta_{k,k'}$. The operator $\Delta B_{k}^\dagger /\sqrt{\Delta t}$ is therefore interpreted as the creation operator for the photons in time-bin $k$, \ie~the photons associated to times included in the interval $[t_k,t_{k+1}[$. 
We thus define the corresponding Fock states $\ket{n}_k\equiv\frac{(\Delta B_k^\dagger)^n}{\sqrt{n!}\sqrt{\Delta t^n}}\ket{\text{vac}}_k$ and write the state of the system at each time-step $k$ as \be \label{eq:defpsi}\ket{\Psi(t_k)}=\sum_{i_S, n_1,n_2,...,n_{N}} \psi_{i_S, n_1, n_2,..., n_N}(t_k) \ket{i_S}_S\otimes\ket{n_1}_1\otimes\ket{n_2}_2\otimes...\ket{n_N}_N, \ee where $i_S$ labels the state of the quantum emitter and $n_j$ denotes the photon number in time-bin $j$. The evolution of the system is then obtained by integrating the QSSE to lowest order in $\Delta t$, which provides the stroboscopic map \mbox{$\ket{\Psi(t_k)}={U_k}\ket{\Psi(t_{k-1})}$} with \be\label{eq:opstrobo}U_k=\text{exp}\Big(- i H_a\,\Delta t +\sqrt{\gamma}\big(\Delta B^\dagger_k \sigma_1 - \Delta B^\dagger_{k-m} \sigma_2 e^{-i\bar\omega\tau} - \text{H.C.}\big)\Big), \ee where the time step interval was chosen such that $\tau=m\Delta t$ with $m$ an integer.

\subsection{Matrix product state ansatz}

The MPS ansatz consists in writing the amplitudes defined in \eqref{eq:defpsi} as products of matrices. In particular, let us express the initial state as 
\be \label{eq:expressmps} \psi_{i_S, n_1, n_2,..., n_N}(t_0) = A[S]^{i_S}\cdot A[1]^{n_1}\cdot A[2]^{n_2}\cdot ... \cdot A[N]^{n_N}.\ee 
Here $A[k]^{n_k}$ is a matrix of dimension $D_k\times D_{k+1}$. Note that $A[S]^{i_S}$ is a vector of dimension $1\times D_1$, and similarly $A[N]^{n_N}$ is of dimension $D_{N}\times 1$. The maximum bond dimension $D\equiv\max_k(D_k)$  determines the maximal amount of entanglement that can be represented with this ansatz.

\subsubsection{Two-photon initial state. ---}
Our MPS approach is naturally formulated in the time domain. To address the two-photon scattering problem, which is formulated in frequency space, we approximate the initial state of two photons (with a given frequency $\bar\omega$) by an incident two-photon wave packet of duration $T=N\Delta t$. This approximation is valid if this duration $T$ is much larger than $1/\gamma$ and $\tau$, hence $N\gg \max(m,1/(\gamma\Delta t))$. 
Denoting the vacuum state in all time-bins by $\ket{\text{vac}}$, we thus consider the two-photon initial state 
\be\begin{aligned} \ket{\Psi(t_0)} &= \frac{1}{\sqrt{2}}\Big(\frac{1}{\sqrt{N \Delta t}}\sum_{k=1}^N\Delta B_k^\dagger\Big)^2\ket{g}_S\otimes\ket{\text{vac}}
\\&= \frac{1}{\sqrt{2} N \Delta t}\Big(\sum_{k=1}^N(\Delta B_k^\dagger)^2 + 2\sum_{k=1}^N\sum_{l=k+1}^N\Delta B_k^\dagger\Delta B_l^\dagger\Big)\ket{g}_S\otimes\ket{\text{vac}}. \end{aligned}\ee 
In order to express this state in the form of \eqref{eq:expressmps} we note that it can be written as
\be \ket{\Psi(t_0)} = \frac{1}{\sqrt{2}N} \sum_{n_1,n_2,...,n_N}\prod_{k=1}^N M[k]^{n_k}\ket{g}_S\otimes\ket{n_1}_1\otimes\ket{n_2}_2\otimes\cdots,\ee
where the matrices $M[k]^{n_k}$ are matrices given by
\begin{align}
M[1]^0=\begin{pmatrix} 1 & 0 & 0 \end{pmatrix}, \ 
M[1]^1=\begin{pmatrix} 0 & \sqrt{2} & 0 \end{pmatrix}, \ 
M[1]^2=\begin{pmatrix} 0 & 0& \sqrt{2} \end{pmatrix} \ 
\end{align}
for $k=1$, 
\be 
M[k]^0=\begin{pmatrix}1 & 0 & 0 \\ 0 & 1 & 0 \\ 0 & 0 & 1 \end{pmatrix}, \ 
M[k]^1=\begin{pmatrix}0 & \sqrt{2} & 0 \\ 0 & 0 & \sqrt{2} \\ 0 & 0 & 0 \end{pmatrix}, \ 
M[k]^2=\begin{pmatrix}0 & 0 & \sqrt{2} \\ 0 & 0 & 0 \\ 0 & 0 & 0 \end{pmatrix}
\ee  
for $1<k<N$, and
\begin{align}
M[N]^0=\begin{pmatrix} 0 &0 &1 \end{pmatrix}^\top,\ 
M[N]^1=\begin{pmatrix} 0 &\sqrt{2} &0 \end{pmatrix}^\top, \ 
M[N]^2=\begin{pmatrix} \sqrt{2} &0 &0 \end{pmatrix}^\top
\end{align}
for $k=N$.
The initial state can thus be represented as an MPS state of the form \eqref{eq:expressmps} with bond dimension $D=3$, with the normalized matrices $A[S]^{i_S}=\delta_{i_S,g}$ and $A[k]^{n_k}= M[k]^{n_k}/(\sqrt{2}N)^{1/N}$ for $1\leq k\leq N$. This expression can be brought to a canonical form in the standard way \cite{Schollwock:2011gl}. 
The time evolution is then performed by applying the unitary operator \eqref{eq:opstrobo} at each time step, hence updating these matrices. For details regarding the implementation, and in particular the long-range interactions arising from the time delay, we refer the reader to \cite{Pichler:2016bx}. We note that it is straightforward to generalize this to states with larger number of photons. 

\subsubsection{Observables. ---}

At the end of the scattering, the MPS state can be analyzed to obtain the properties of the output field. The first order correlation function reads
\be \label{eq:obsG1} G^{(1)}(t_{k+m}, t_k)=\langle b^\dagger(t_{k+l})b(t_k)\rangle\approx\langle\Delta B^\dagger_{k+l}\Delta B_{k}\rangle/(\Delta t)^2, \ee and the inelastic output spectrum is given by 
\be S_\text{inel}(\nu)=\frac{1}{\pi}\int_0^{\infty}dt' \text{Re}\Big(\tilde G^{(1)}(t',t_k)e^{i\nu t'}\Big)\approx \frac{\Delta t}{\pi}\sum_{l=0}^M \text{Re}\Big(\tilde G^{(1)}(t_{k+l},t_k)e^{i\nu l \Delta t }), \ee
where $k$ and $M$ should be taken large enough so that the system is in the steady-state and $\tilde G^{(1)}$ is obtained from $G^{(1)}$ by subtracting the time-independent offset corresponding to the elastic part of the spectrum.
The normalized second order auto-correlation function on the other hand reads 
\be \label{eq:obsg2}g^{(2)}(t_{k+l},t_k)=\frac{\langle b^\dagger(t_k)b^\dagger(t_{k+l})b(t_{k+l})b(t_k)\rangle}{\langle b^\dagger(t_k)b(t_k)\rangle\langle b^\dagger(t_{k+l})b(t_{k+l})\rangle}\approx\frac{\langle\Delta B_k^\dagger\Delta B_{k+l}^\dagger\Delta B_{k+l}\Delta B_k\rangle}{\langle\Delta B_k^\dagger\Delta B_k\rangle\langle\Delta B_{k+l}^\dagger\Delta B_{k+l}\rangle}.\ee
These correlation functions of the output state can be calculated using standard MPS routines \cite{Schollwock:2011gl}.
In our simulations presented below we calculate these observables using the MPS parameters $\gamma\Delta t =0.1$, $D=200$ and $\gamma T=300$.

\subsubsection{Coherent initial state. ---}
Within the MPS formalism coherent input fields can be naturally included by a Mollow transformation \cite{QWII} where the noise increments are redefined as \mbox{$\Delta B_k  = \int_{t_{k-1}}^{t_k} dt\, b(t) - \beta\, \Delta t$} for a finite photon intensity of $|\beta|^2$ (with $\beta\equiv\alpha/\sqrt{T})$. Correspondingly the unitary operator of \eqref{eq:opstrobo} that induces the evolution now reads
\be \begin{aligned}U_k=\text{exp}\Big(-i H_a\,\Delta t  &+\big(\Omega_R(\sigma_1 - \sigma_2 e^{-i\bar\omega\tau}) - \text{H.C.}\big) \Delta t\\&+\sqrt{\gamma}\big(\Delta B^\dagger_k \sigma_1 - \Delta B^\dagger_{k-m} \sigma_2 e^{-i\bar\omega\tau} - \text{H.C.}\big)\Big). \end{aligned}\ee
The additional terms that appear under this transformation act as a classical driving field with Rabi frequency $\Omega_R\equiv\beta^*\sqrt{\gamma}$ on the atomic degrees of freedom. Moreover the input state of the waveguide in this transformed picture is simply the vacuum state. In a MPS description it is trivially represented by $A[j]^{n_j} = \delta_{n_j,0}.$
The observables \eqref{eq:obsG1} and \eqref{eq:obsg2} are then redefined with $\Delta B_k\to\Delta B_k+\beta\, \Delta t$.  With this approach arbitrarily strong coherent driving fields can be handled numerically \cite{Pichler:2016bx}.

\section{Results}
\label{sec:Results}
{In the previous sections we developed an analytical solution and a numerical method to tackle the scattering problem.}
In this section we display and discuss the results obtained using these two approaches. We start out with presenting the properties of the field scattered by a two-level emitter, and then discuss the similarities and differences to the case of a chiral V-level emitter. 
As expected, the numerical calculations and the analytical solutions show a perfect agreement in the 2-photon sector.

\subsection{First order correlation function: inelastic spectrum}
\begin{figure}
\begin{center}
\includegraphics[width=\textwidth]{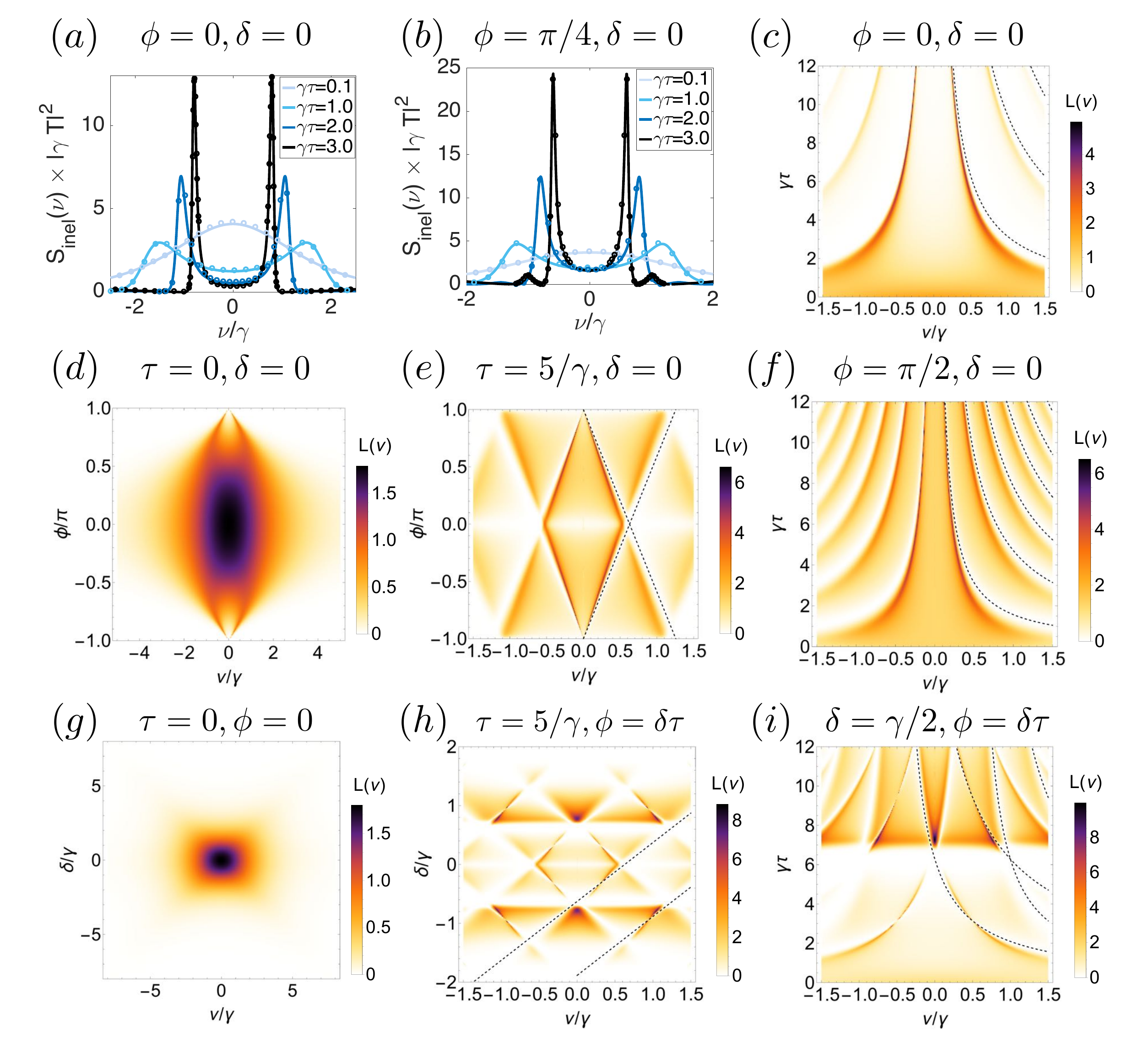}
\caption{\label{fig2} Inelastic spectrum of the 2-photon state scattered by a two-level quantum emitter. We consider various values of the coherent quantum feedback delay $\tau$ with $\delta=0$, and $\phi\equiv\bar\omega\tau+\pi=0$ in (a) and $\phi=\pi/4$ in (b). The solid lines are obtained using the MPS approach, and we compare the results with the scattering theory approach, in circles. In (c)-(i) we plot the logarithm of the spectrum $L(\nu)\equiv\log(1+S_\text{inel}(\nu)|\gamma T|^2)$ calculated via the scattering theory approach. The various expressions for the dashed black lines are given in the text.}
\end{center}
\end{figure}
The inelastic spectrum of the output field is sensitive to processes in which the photons change frequency during the scattering. Due to energy conservation this can happen only if more than one photon is involved in the scattering process. 
For the case of two incident photons of the same frequency considered here, this leads to a pair of photons with opposite frequencies in the scattered field. The inelastic spectrum is therefore necessarily  symmetric around $\nu=0$. 
In Figs.~\ref{fig2}(a)-(b) we show examples of the inelastic spectra for different delay times. To understand the basic physics we  discuss the two limiting cases of the Markovian regime and the non-Markovian regime separately.
\subsubsection{Markovian regime ---}
If the delay time is much smaller than the inverse bandwidth of the transition, \ie~$\gamma\tau\ll1$, the system is effectively Markovian, as the delay line is too short to introduce significant memory. In this regime the system is effectively described by a master equation, where the mirror only renormalizes the transition frequency of the two-level system and its coupling to the field, as $\delta_\text{eff}=\delta-\gamma \sin(\phi)$ and $\gamma_\text{eff}=4\gamma\cos^2(\phi/2)$ \cite{Dorner:2002dv}. These effective parameters essentially depend on the distance of the atom from the mirror, via the propagation phase $\phi$. For $\phi=\pi$ the incident field interferes destructively with the field reflected from the mirror and the atom decouples from the waveguide. On the other hand, for $\phi=0$ the interference is constructive, leading to an enhanced emission of photons. The renormalization of the transition frequency can be interpreted as a dipole interaction of the atom with its mirror image. 

The scattering problem of a Markovian two-level system is well understood, and can be calculated via the optical Bloch equations. From the general formula \eqref{Sel_inel} we recover the expression for the inelastic spectrum in the Markovian limit:
\begin{align}
S_{\rm inel}\xrightarrow{\gamma\tau\rightarrow 0} \frac{64}{\pi (\gamma T)^2}\frac{\gamma_{\rm eff}^2/4}{\delta_{\rm eff}^2+\gamma_{\rm eff}^2/4}\frac{\gamma_{\rm eff}^2/4}{\left((\nu+\delta_{\rm eff})^2+\gamma_{\rm eff}^2/4\right)\left((\nu-\delta_{\rm eff})^2+\gamma_{\rm eff}^2/4\right)}.
\end{align}
If the incident photons are on resonance with the atom ($\delta=0$) the spectrum can have one or two peaks, depending on $\phi$, (see Fig.~\ref{fig2}(d)). This depends on whether the effective level shift introduced by the mirror $\delta_{\rm eff}$ is resolved on the scale of the effective decay rate $\gamma_{\rm eff}$. For $\phi\approx 0$ the spectrum has a single maximum at $\nu=0$ and is dominated by the broad effective decay rate $\gamma_{\rm eff}$. Away from $\phi=0$, the central peak becomes accordingly more narrow, until eventually one can resolve two peaks at $\nu=\pm \delta_{\rm eff}$. These two peaks stem from the fact that the feedback shifts the atom out of resonance: incoherent scattering into photons resonant with the renormalized frequency is enhanced. That is, one of the two incoherently scattered photons is then resonant with the renormalized atomic transition frequency. 
If the atom is placed exactly at $\phi=\pi$ the coupling to the waveguide and correspondingly the inelastic spectrum vanishes. 
Finally, if the photons are off resonance with the atom, the results have a similar interpretation (see Fig.~\ref{fig2}(g)): for a finite value of the detuning $\delta$ one resolves peaks at $\nu=\pm\delta$, albeit with lower amplitude than in the resonant case.

\subsubsection{Non-Markovian regime ---}
In the non-Markovian regime $\gamma\tau\gg1$ the inelastic spectrum is much richer. In the simplest case of $\phi=0$, $\delta=0$ it can be expressed as 
\be S_\text{inel}(\nu)=\Big(\frac{8}{\sqrt{\pi}\gamma T (1+\gamma\tau)}\frac{1+\cos(\nu\tau)}{(\nu/\gamma-\sin(\nu\tau))^2+(1+\cos(\nu\tau))^2} \Big)^2.\ee  and displays a prominent pair of sharp peaks (see Fig.~\ref{fig2}(a)) and an infinite number of smaller side peaks (see Fig.~\ref{fig2}(c)). 
To understand the basic physics in this non-Markovian limit we can interpret this systems as a leaky cavity formed by the real perfect mirror on one side and the two-level atom playing the role of an imperfect mirror on the other side \cite{Chang:2012wb,PhysRevA.93.023808}. The resonance frequencies and the effective linewidth of this cavity depend on the delay time and the propagation phase. 
For $\phi=0$ and $\delta=0$, the atom is completely decoupled from photons with frequencies $\nu$ given by an odd multiple of $\pi/\tau$: indeed, all of these photons pick up a phase $\pi$ when propagating to the mirror and back, such that their contribution to the electric field vanishes at the atomic position due to destructive interference. 
This is shown in Fig.~\ref{fig2}(c), where the dashed lines display these frequencies as a function of $\gamma\tau$. Along these lines the inelastic spectrum vanishes, since the atom cannot scatter incoming photons into these modes. These frequencies however also coincide with the frequency of the modes supported by the effective cavity. The resulting effect is the formation of peaks in the intensity of the scattered photons very close to these zeros, corresponding to cavity resonances. As the delay increases, the peaks thus get closer to $\nu=0$, and get sharper, corresponding to a decrease of the effective cavity linewidth. This is due to the fact that the effective cavity mode frequency approaches the two-level resonance, which improves the efficiency of the atom as a mirror, leading to an effective high finesse cavity.

Similarly, for a general $\phi$ the propagation phase acquired by a photon of frequency $\nu$ propagating from the atom to the mirror and back is given by $\phi+\nu\tau$. Thus, the atom decouples from photons with frequencies given by $((2l-1)\pi+\phi)/\tau$, with $l$ an integer. In Figs.~\ref{fig2}(e),(f) this is visible as the vanishing of $S_{\rm inel}$ along these curves, represented in dashed lines. Due to the symmetry of the spectrum, it must also vanish at $\nu=((2l-1)\pi - \phi)/\tau$. This decoupling comes again hand in hand with resonances of the effective cavity close to these frequencies and we observe the corresponding sharp peaks of the inelastic spectrum.

Finally, for a finite value of the detuning $\delta$ between photons and atom we can apply a similar reasoning. In order for the modes supported by the effective cavity to be independent of the incoming photon detuning, in Figs.~\ref{fig2}(h),(i) we choose to set the atomic phase $\omega_a\tau+\pi$ to $0$ (modulo $2\pi$), rather than $\phi$, which then reads $\phi=\delta\tau$. The frequencies of the decoupled photons are now shifted by $\pm\delta$, as is shown in Figs.~\ref{fig2}(h),(i) where the dashed lines corresponds to $\nu=(2l-1)\pi/\tau \pm\delta$ (with $l$ an integer). Further we notice that the spectrum identically vanishes whenever $\phi=\delta\tau$ is an odd multiple of $\pi$ (see Fig.~\ref{fig2}(i)), as expected.

\subsection{Second order correlation function}
To understand the temporal statistics of the two photons, we analyse the second order auto-correlation function $g^{(2)}(t)$. For the two-photon input state the correlation function is flat and evaluates to $g^{(2)}=1/2$, \ie~there are no correlations (apart from those given by the fixed total photon number). The atom as a non-linear element can induce non-trivial correlations between photons as a result of the scattering process. 

 \begin{figure}
\begin{center}
\includegraphics[width=\textwidth]{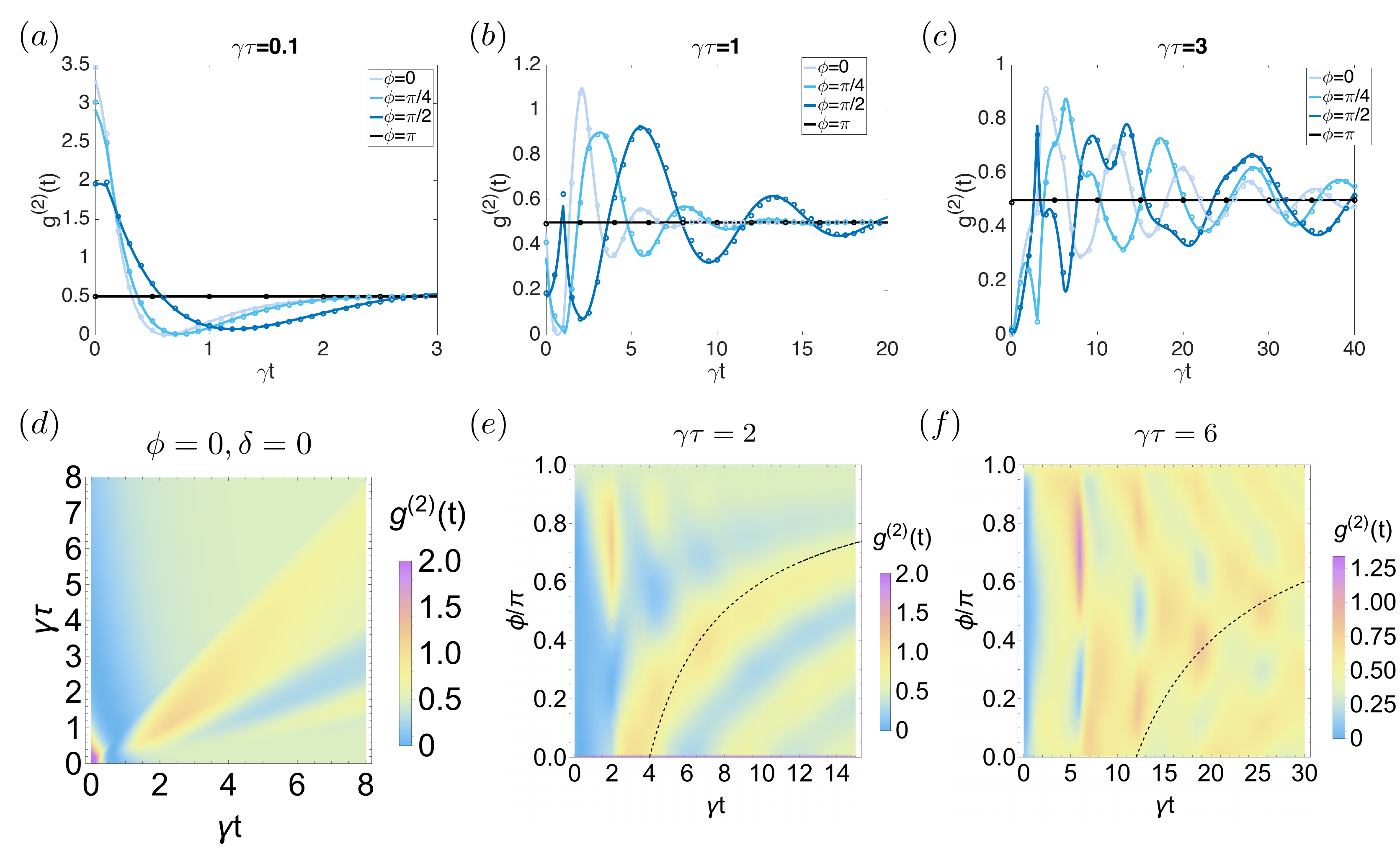}
\caption{\label{fig3} Second order correlation of the 2-photon state scattered by a two-level quantum emitter with $\delta=0$. In (a)-(c) the solid lines are obtained using the MPS approach, and we compare the results with the scattering theory approach, in circles. In (d)-(f) we use the scattering theory approach. The dashed black lines in (e),(f) are given in the text.}
\end{center}
\end{figure}

Again, we first discuss the results in the Markovian limit $\gamma\tau\ll1$, shown in Fig.~\ref{fig3}(a), where the entire setup can be understood as an effective two-level system as described above. In this limit the output field exhibits bunching of the photons at equal times, $g^{(2)}(t=0)>1/2$. This can be understood as follows: once the first photon is absorbed the second photon leads to an elastic stimulated emission. This increases the probability for observing the two photons at the same time at the output port of the waveguide, as signaled by $g^{(2)}(t=0)>1/2$. The same effect leads to a decrease of the probability to detect two photons separated by $t\sim1/\gamma$, as visible in Fig.~\ref{fig3}(a): the fast stimulated emission of the first absorbed photon by the second photon depletes the output field at later times (with respect to the uncorrelated state) leading to an almost perfect anti-bunching dip. Note that again the point $\phi=\pi$ is special: the atom decouples from the photons in this case and accordingly photons stay uncorrelated. 

Remarkably, when one increases the time delay and leaves the Markovian limit, the auto-correlation function becomes non-analytic at each integer multiple of the delay time $t=n\tau$, $n=1,2,3,\dots$. This is most prominently visible at $t=\tau$, where the first derivative of $g^{(2)}(t)$ is dicontinuous (see Figs.~\ref{fig3}(b)-(d)). These non-analyticities arise from the fact that a photon can ``bounce'' multiple times between the atom and the mirror.
We note that these non-analyticities are always present for any nonzero $\tau$, however they become less pronounced for short delay times $\tau\rightarrow 0$, such that $g^{(2)}$ eventually approaches a continuous function, as predicted in the Markovian approximation. 

One of the most striking features of the auto-correlation function deep in the non-Markovian regime are ``long-ranging'' correlations between the photons as indicated by $g^{(2)}(t)\neq 1/2$ even for $t\gg 1/\gamma$. In particular, for $t\gg\tau$, $g^{(2)}(t)$ displays damped oscillations, where the oscillation frequency and their decay rate depend on $\phi$ (for simplicity we consider here  only $\delta =0$). To understand this behavior we recall that the output state for $\gamma\tau\gg1$ is a superposition of the elastically scattered photons, and a state containing two photons at opposite frequency $\pm \nu$. From the discussion in the previous section we know that the distribution of the frequencies of these incoherently scattered photons, $\hat t(\nu)$, is strongly peaked around $\nu=\pm(\pi-|\phi|)/\tau$, corresponding to the supported frequency of the cavity formed by the atom and the distant mirror (see Fig.~\ref{fig2}(e)). From Eq.~\eqref{g2def} it is clear that the auto-correlation function in such a state exhibits oscillations $\sim \cos(\nu t)$, \ie~$g^{(2)}$ is approximately periodic with period $\frac{2\pi\tau}{\pi-|\phi|}$. In Figs.~\ref{fig3}(e),(f), we see that the oscillation period follows well this predicted curve as a function of $\phi$, plotted in dashed lines. For $\phi\rightarrow \pi$ this period diverges, but at the same time the amplitude of the oscillations vanishes, such that $g^{(2)}(t)=1/2$, \ie~the atom decouples from the waveguide as in the Markovian case. 

The decay of the oscillations in the auto-correlation function is directly related to the finite linewidth of the effective cavity modes supported by the mirror and the atom. This is also consistent with the expectation that correlations between the two photons in the output field can not extend over longer times than the timescale over which the effective cavity can store a photon and in this way introduces a long memory time. Indeed the reflectivity of the atom as a mirror reads \cite{Chang:2012wb,PhysRevA.93.023808} $R=1/(1+(\nu/\gamma)^2)$ for a frequency mismatch $\nu$ between the atomic transition and the photon. We can thus estimate the lifetime for a single photon inside the cavity in a mode with frequency $\nu$ as $t_\text{eff}\sim \tau(1-R)\sum_{n=0}^\infty(n+1)R^n=\frac{\tau}{1-R}=\tau(1+(\frac\gamma{\nu})^2).$ Here each term of the sum corresponds to the probability for a photon to be reflected $n$ times (with probability $R$), before leaving the system (with probability $1-R$). The fact that the correlations extend over long times, is thus a direct consequence of the sharpness of the resonance peaks in the inelastic spectrum close to $\nu=0$. As the mode frequency reads $\nu=\pm(\pi-|\phi|)/\tau$, we conclude that the oscillation decays slowly for large $\tau$ or for $\phi$ close to $\pi$, as can be observed in Fig.~\ref{fig3}.

\subsection{V-level system}
\begin{figure}
\begin{center}
\includegraphics[width=\textwidth]{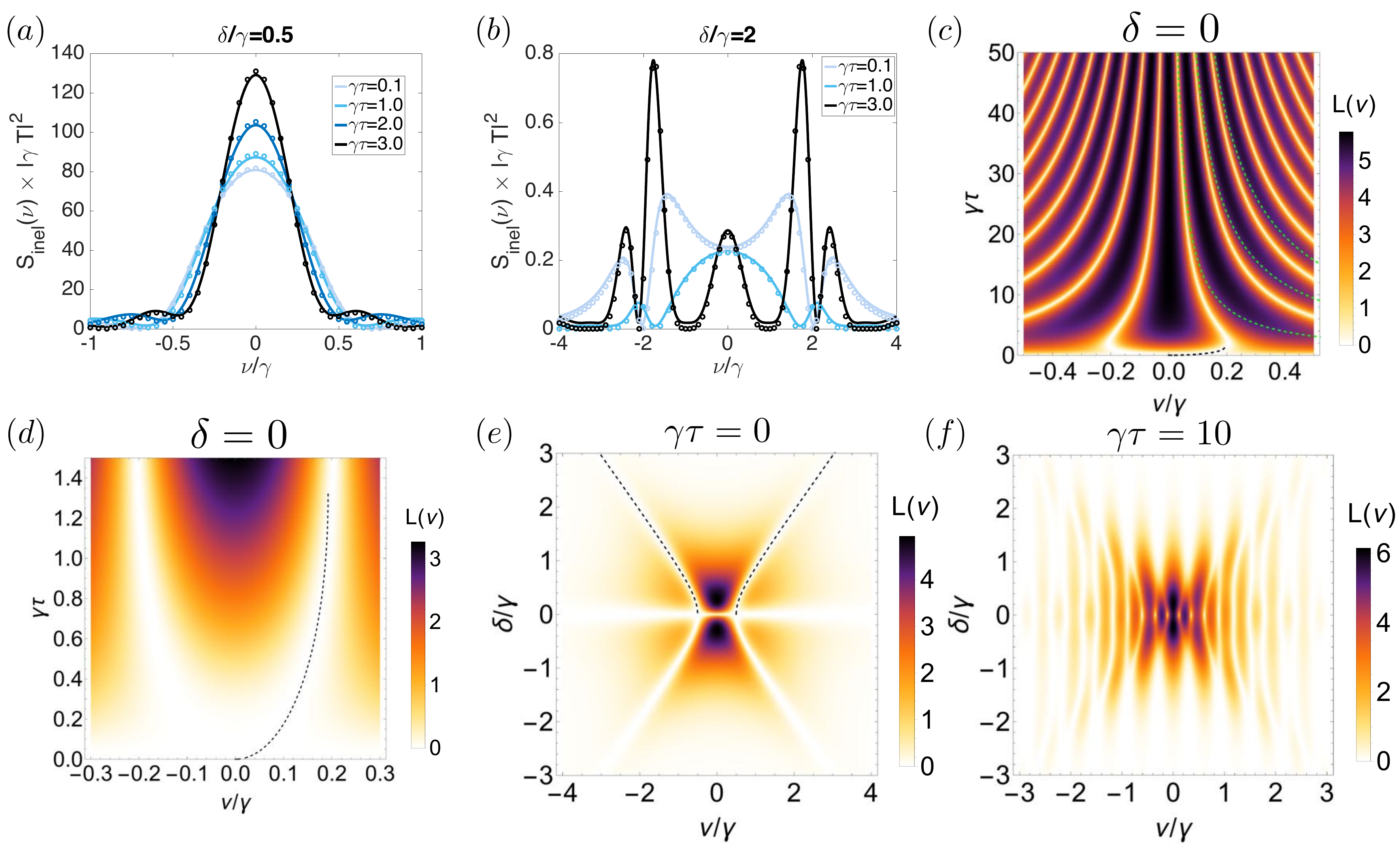}
\caption{\label{fig4} Inelastic spectrum of the 2-photon state scattered by a V-level quantum emitter. In (a) and (b), the solid lines are obtained using the MPS approach, and we compare the results with the scattering theory approach, in circles. In (c)-(f) we plot the logarithm of the spectrum $L(\nu)\equiv\log(1+S_\text{inel}(\nu)|\gamma T|^2)$ using the scattering theory approach. The dashed lines are given in the text. }
\end{center}
\end{figure}
If we replace the two-level scatterer by a chiral V-level system the situation changes both in the Markovian and in the non-Markovian regime (see Figs.~\ref{fig4} and \ref{fig5}). First we note that the propagation phase $\phi$ picked up by the incident photons when traveling from the atom to the mirror and back can be gauged away in this setting and thus has no measurable physical consequences. A second crucial difference to the previous case is that due to the chiral interactions, photons can only be scattered forward by the atom. Therefore each photon can only interact with the atom twice. As a consequence the atom can not form an effective cavity with the mirror, such that none of the above discussion can be transferred to the case of a chiral V-level scatterer. For simplicity we limit the discussion in what follows to the case of equal detunings for the two transitions $\delta_1=\delta_2\equiv\delta$.

\subsubsection{Markovian regime ---}
First, in the Markovian limit ($\gamma\tau\rightarrow 0^+$) the system can again be described by a master equation. In this case the mirror however does not only renormalize the coupling to the photon field and the transition frequencies, but it also induces an effective coherent coupling between the two excited states, as was shown in \cite{Guimond:2016kf}. In particular for $\delta=0$ there exists a dark state such that the atom completely decouples from the photon field, in which case the inelastic spectrum identically vanishes (see Figs.~\ref{fig4}(d),(e)). Thus in the Markovian regime nontrivial features in the scattered field can only be obtained with a finite detuning $\delta$. In Fig.~\ref{fig4}(e), we show the inelastic spectrum in this case. One would expect an enhanced scattering into photons with resonant frequencies $\nu=\pm\delta$. However, most strikingly the atom here decouples from photons at frequencies satisfying $(\nu/\gamma)^2+(\delta/\gamma)^2=1/4$ as plotted in dashed lines, which is due to a destructive interference effect. For $\delta\gtrsim\gamma/2$ this results into a splitting of the two resonant peaks $\nu=\pm\delta$ into four, as can be observed in the examples of Figs.~\ref{fig4}(a),(b).

\subsubsection{Non-Markovian regime ---}
If one increases the delay $\tau$, the scattered field is non-trivial even at $\delta=0$. The expression of the inelastic spectrum for $\delta=0$ now reads \be S_\text{inel}(\nu)=\frac{4 e^{-2 \gamma\tau } (4\gamma)^4 (e^{\gamma\tau /2}-1)^2}{\pi(\gamma T)^2  \left(4 \nu^2+\gamma^2\right)^4}  \left[4\nu^2+\gamma^2+e^{\gamma\tau /2} \left((4 \nu^2-\gamma^2) \cos (\nu \tau )+4 \nu\gamma \sin (\nu  \tau )\right)\right]^2. \ee
For small values of the delay an expansion in $\gamma\tau$ provides an expression for the zeros of the spectrum at $\nu/\gamma=\frac{1}{4}\sqrt{\gamma\tau }-\frac{1}{16}(\gamma\tau)^{3/2}+\mathcal O(\gamma\tau)^{5/2}$, as pictured in black dashed lines in Figs.~\ref{fig4}(c),(d).
For larger delays the spectrum is dominated by the non-Markovian contributions. Similar to the case of the two-level system it displays a series of peaks separated by points where the spectrum vanishes, which display a $~1/\tau$ dependency. This effect arises here because of interference between photons scattered by the different atomic transitions. More precisely, in an inelastic event pairs of photons with frequencies $\pm\nu$ can be produced. Let us consider an event where the photon with frequency $\nu$ is generated in a scattering event from transition $1$ while the photon with frequency $-\nu$ is created by transition $2$. The amplitude of this event acquires a phase $e^{i\nu\tau}$ due to the propagation of the first photon in the delay line. Conversely, the amplitude for the event with opposite frequencies acquires a phase $e^{-i\nu\tau}$, hence, for $\nu\tau=\pi/2$ (modulo $\pi$) these contributions cancel, and correspondingly no photons are created at these frequencies. The dashed green lines in Fig.~\ref{fig4}(c) display these frequencies as a function of $\tau$, which shows agreement in the limit of large delays.

Finally, for finite values of the detuning $\delta$, we observe a rich variety of peaks in the inelastic spectrum (see Fig.~\ref{fig4}(f)) which strongly depend on $\delta$, similarly to the Markovian case (Fig.~\ref{fig4}(e)). However the density of peaks is now much higher due to the behavior of their frequencies in $1/\tau$. Again, for small values of $\delta\lesssim\gamma/2$ the central peak at $\nu=0$ is dominant (see also Fig.~\ref{fig4}(a)), while for $\delta\gtrsim\gamma/2$ the stronger peaks are located close to $\nu=\pm\delta$ (see also Fig.~\ref{fig4}(b)).

\subsubsection{Second order correlation function ---}

\begin{figure}
\begin{center}
\includegraphics[width=12cm]{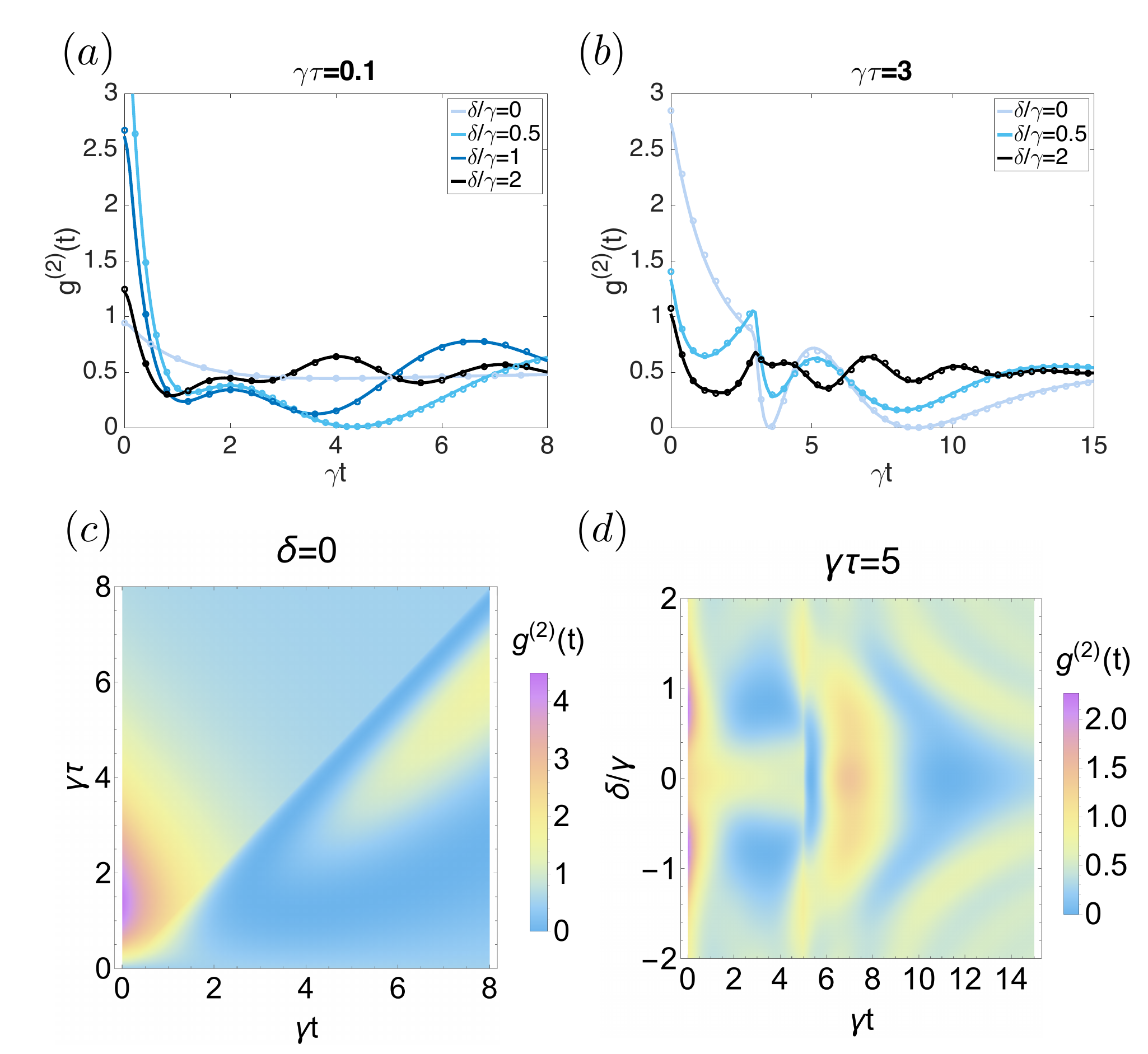}
\caption{\label{fig5} Second order correlation of the 2-photon state scattered by  a V-level quantum emitter. In (a) and (b) the solid lines are obtained using the MPS approach, and we compare the results with the scattering theory approach, in circles. In (c) and (d) we use the scattering theory approach.}
\end{center}
\end{figure}

We now analyse the temporal statistics of the photons in the second order correlation function for the case of a V-level emitter. In the Markovian regime (see Fig.~\ref{fig5}(a)), the situation is similar to the case of a two-level emitter, and the photons are bunched (\ie~$g^{(2)}(0)>1/2$) as an effect of stimulated emission. Here we additionally show results for a finite detuning $\delta$, which generates small decaying oscillations. Again, from Eq.~\eqref{g2def} the oscillation frequencies are given by the positions of the most prominent peaks in the inelastic spectrum (Fig.~\ref{fig4}(e)). The period of the oscillations is thus approximately given here by $2\pi/\delta$. 

In the non-Markovian regime however, the system displays different results with respect to the response of a two-level emitter. Indeed, due to the chiral interactions the scattered photons cannot bounce back indefinitely between atom and mirror, but rather interact only once with each atomic transition. Here $g^{(2)}(t)$ is thus non-analytic only at $t=\tau$ (see Fig.~\ref{fig5}(b)), as the absence of an effective cavity forbids the storage of a photon for long times and consequently prevents long lasting oscillations in $g^{(2)}$. In order to gain more insight in the shape of the correlation function, let us first assume $\delta=0$ (see Fig.~\ref{fig5}(c)). The non-analyticity here displays a strong contrast only along the line $t=\tau$.  When the delay increases, the bunching vanishes as $g^{(2)}(0)\to1/2$, indicating that photons are scattered mostly by different transitions. This can be further observed as there is a strong correlation in $g^{(2)}$ right after time $t=\tau$ for large values of the delay.
Finally, for a finite detuning we observe, on top of these features, oscillations, once again with a period approximately given by $2\pi/\delta$. While with $\delta=0$ the auto-correlation $g^{(2)}(t)$ converges towards $1/2$ for $t<\tau$ in Fig.~\ref{fig5}(d), a value of $\delta\approx\gamma$ generates richer correlations as $g^{(2)}(t)$ undergoes oscillations with a large amplitude. The decrease of the oscillation period as $1/\delta$ is there clearly visible for $t>\tau$. 

\subsection{Finite coherent driving}
\begin{figure}
\begin{center}
\includegraphics[width=\textwidth]{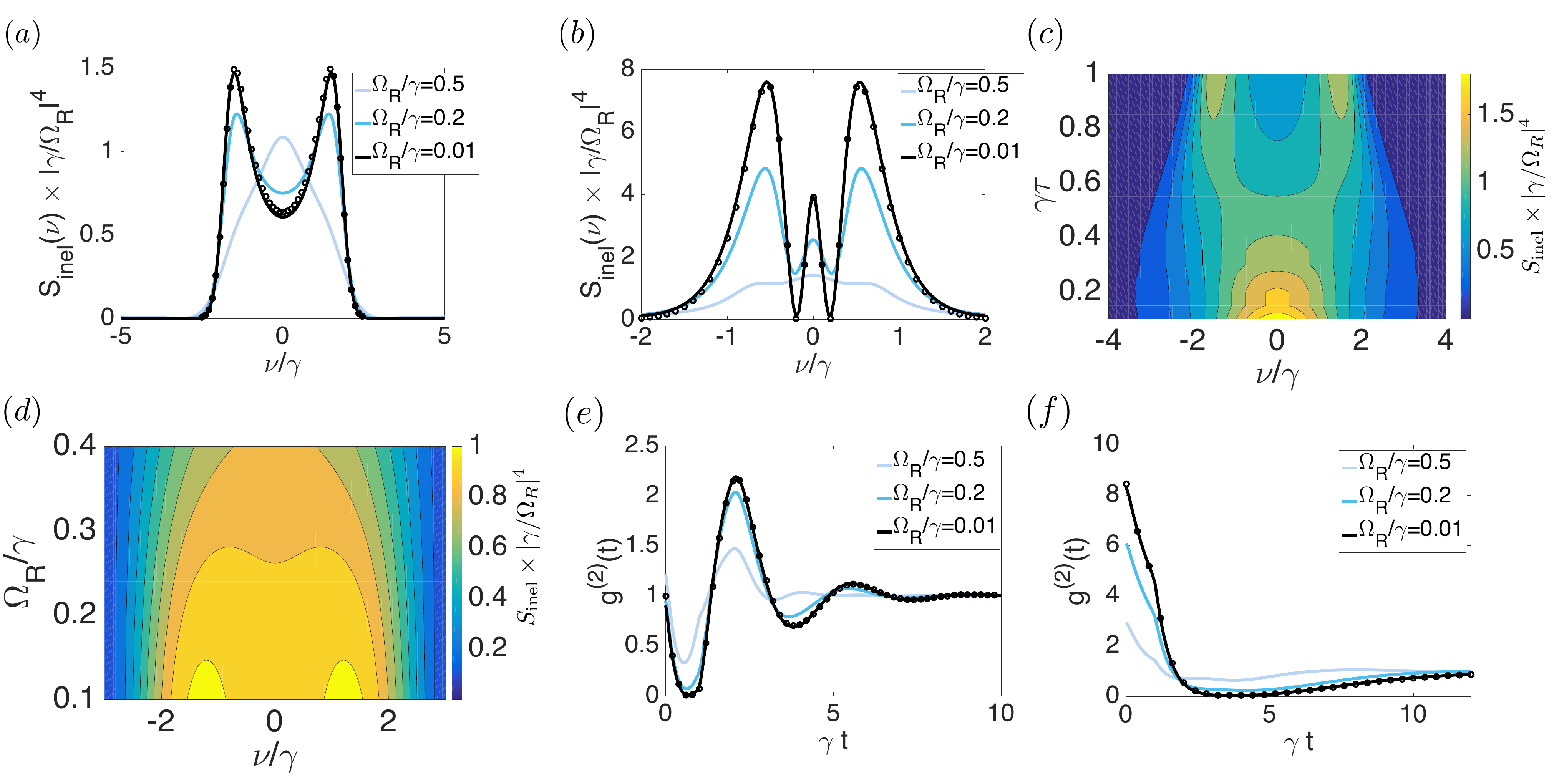}
\caption{\label{fig6} (a)-(d) Inelastic spectrum and (e),(f) second order correlation of the output field of a coherently driven quantum emitter, with $\phi=0$ and $\delta=0$. In (a),(c)-(e) we consider a two-level emitter, and in (b),(f) a V-level emitter. In (a),(b),(e),(f), the solid lines are obtained using the MPS approach with $\gamma\tau=1$, and are compared with the predictions of the scattering theory approach in the weak field limit, in black circles. In (c),(d) we use the MPS approach with (c) $\Omega_R/\gamma=0.1$ and (d) $\gamma\tau=0.6$.}
\end{center}
\end{figure}
 Up to now we have analysed only the scattering of two-photon states, which in theory have an infinite spread and thus a vanishing intensity.
In this final section we conclude our analysis by considering, on the other hand, the effect of a {\it finite} driving, that is, the scattering of coherent states with a nonzero photon intensity. In Fig.~\ref{fig6} we calculate the spectrum and the second order correlation function for different values of the Rabi frequency $\Omega_R$ of the driven atom, in the non-Markovian regime. We see that the analytical formulas, based on a truncation of the state of the driving field to few-photon components, agree well with the quasi-exact numerical results in the limit $\Omega_R\rightarrow 0$. On the other hand, for finite coherent driving intensities, contributions from components with more than two photons become more and more significant. 
We observe in Figs.~\ref{fig6}(a),(b) that the inelastic portion of the spectrum decreases for strong driving. This can be understood as the saturation of the atom, which prevents most of the driving field to be inelastically scattered. For the same reason, the second order auto-correlation function, shown here in Figs.~\ref{fig6}(e),(f), identically convergences towards $1$, corresponding to a perfectly coherent output. 

As we showed in the previous sections, the physics underlying the results is vastly different between Markovian and non-Markovian regimes. Here the transition between these two regimes is displayed in Fig.~\ref{fig6}(c) for a fixed driving intensity, showing at large delays the presence of the peaks characteristic of the first modes of the cavity formed by the mirror and the atom. However, in Figs.~\ref{fig6}(a),(b) we also observe that by increasing the driving intensity, the shape of the spectrum converges towards the usual Lorentzian centred around $\nu=0$, as in the Markovian regime. This can be understood as the saturation of the system preventing the existence of all interference effects discussed in the previous sections. This effect is further shown in Fig.~\ref{fig6}(d) for a fixed value of the time delay. This shows that the threshold for the transition between Markovian and non-Markovian behavior depends on the competition between time delay and driving intensity. 

\section{Conclusion}

In this work we have studied the scattering problem in a setup with delayed coherent quantum feedback, where light propagating in a semi-infinite waveguide scatters off of an atom located in front of a mirror. In particular we have analyzed the inelastic scattering of two photons and shown how the output spectrum strongly depends on the delay of the coherent feedback. We used a scattering theory approach, which provides analytical insight and is suitable for few-photon problems, and a numerical method based on matrix product states, which is designed to simulate complex problems involving larger numbers of photons. While these methods are tailored for different questions, their predictions can be compared and have shown strong agreement in the weak driving limit.

\section*{Acknowledgements}
{P.O.G., H.P. and P.Z. provided the numerical method, while M.P. contributed the analytical approach.}
Work at Innsbruck is supported by SFB FOQUS and ERC Synergy Grant UQUAM. H.P. is supported by the NSF through a grant for the Institute for Theoretical Atomic, Molecular, and Optical Physics at Harvard University and the Smithsonian Astrophysical Observatory.

\appendix

\section{Explicit analytic expressions}
\label{app:explicit}
Here we provide expressions for the variables defined in section \ref{sec:diagrammatic} for the two types of quantum emitter we consider.

\subsection{Two-level atom}

Let us first consider the atom as a two-level system, as is represented in Fig.~\ref{fig1}(a). If we define \mbox{$\sigma\equiv \ket{g}\bra{e}=\sigma_1=\sigma_2$}, the self-energy in \eqref{eq:defSigma} reads \mbox{$\Sigma(E)= - i \gamma (1-e^{i(\bar\omega+E) \tau}) \ket{e}\bra{e}$}, and the atomic dressed Green's function in the excited state from \eqref{eq:dyson} becomes
$M(E) = \ket{e}\bra{e}(E + \lambda -  i \gamma e^{i(\bar\omega+E) \tau})^{-1}$,
where $\lambda = \delta +  i \gamma$, with $\delta$ the atomic detuning. The scattering phase in \eqref{eq:stnudef} then expresses as \be s =(\lambda^* +  i \gamma  e^{-i \bar\omega\tau})/(\lambda -  i \gamma  e^{i \bar\omega \tau}).\ee From the definition in \eqref{eq:defOmega}, we also obtain 
\be
\bra{\nu'}\Omega(E)\ket{\nu} = \ket{e}\bra{e}\frac{\gamma}{2 \pi} \frac{(e^{i (\bar \omega + \nu)\tau/2} - e^{-i (\bar\omega+\nu)\tau/2} )  (e^{-i (\bar \omega + \nu')\tau/2} - e^{i (\bar \omega+\nu')\tau/2} ) }{E-\nu' - \nu + i \eta}. 
\label{wtls}
\ee

In order to obtain an expression for $t(\nu)$ in \eqref{eq:stnudef}, we first solve for $W(E)$ in the recursive equation \eqref{eq:recWE}. Details of the method can be found in the Suppl. Mat. of \cite{Laakso:2014bk}. The solution can be expressed as
\be
\bra{\nu'}W(E)\ket{\nu} =  \ket{e}\bra{e}\frac{\gamma}{2 \pi} (e^{i (\bar \omega + \nu)\tau/2} - e^{-i (\bar\omega+\nu)\tau/2} )  (e^{-i (\bar \omega + \nu')\tau/2} - e^{i (\bar \omega+\nu')\tau/2} )\overline{W}_{\nu',\nu} (E),
\ee
where $\overline W_{\nu',\nu} (E)$ satisfies
\be
\overline{W}_{\nu',\nu} (E) = \frac{1}{E - \nu' - \nu + i \eta} - \frac{\gamma}{2 \pi} \int d \nu_1  \frac{1}{E - \nu' - \nu_1 + i \eta} \frac{e^{i (\bar \omega+\nu_1)\tau}}{E - \nu_1 + \lambda} \overline{W}_{\nu_1 ,\nu} (E) .
\label{Wred}
\ee
Note that the first term corresponds to the Markovian contribution, whereas the second term is non-negligible only when $\gamma\gtrsim 1/\tau$. For an input state with $E=0$, we then obtain (see also \cite{PhysRevA.95.053821})
\be
\overline{W}_{-\nu, 0} (0) = \frac{1}{\nu + i \eta} + \frac{i \gamma e^{i\bar \omega\tau}}{\lambda - i \gamma e^{i \bar\omega\tau}} \sum_{\sigma= \pm , 0} C_{\sigma} \frac{e^{i \nu \tau}- e^{-i \sigma p \tau}}{\nu + \sigma p},
\ee
where $p \equiv \sqrt{\lambda^2 + \gamma^2 e^{2 i \bar\omega\tau}}$, and the coefficients are defined as \mbox{$C_0 = -1$} and 
\mbox{$C_{\pm} = \pm ({[\pm p - \lambda] e^{\pm i p \tau} +  i \gamma e^{i \bar\omega\tau}})/({2p \cos p\tau - 2i \lambda \sin p\tau})$}.

In terms of this solution we finally evaluate the output power spectra in \eqref{Sel_inel} as
\begin{align}
S_\text{el}(\nu)= &\left[\frac{2 c }{L}- \frac{4c^2}{L^2} 16\, \gamma^2 \big(1-\cos \bar\omega \tau\big)^2  \mathrm{Im} \big(  [s^*]^2  \Lambda\, m^3 (0) \big)  \right] \delta (\nu), \\
S_\text{inel}(\nu)= &\frac{(4c)^2}{L^2\pi}  \gamma^4 (1- \cos \bar\omega\tau)^2 |m (0)|^2 (\cos\nu\tau-\cos \bar\omega\tau)^2 \nonumber \\
& \times  \bigg|\frac{m (\nu ) - m (-\nu)}{\nu} + m (\nu ) F (\nu) + m (-\nu ) F (-\nu) \bigg|^2,
\end{align}
where we defined the variables 
\begin{align}
F (\nu) \equiv& -({i e^{-i\bar\omega\tau}\lambda/\gamma +1})^{-1} \sum_{\sigma= \pm , 0} C_{\sigma} ({e^{i \nu \tau}- e^{-i \sigma p \tau}})/({\nu + \sigma p}),\\
m (\nu) \equiv& ({\nu + \lambda - i \gamma e^{i (\bar\omega+\nu) \tau}})^{-1},\\
\Lambda \equiv& 1 - {i \gamma e^{i\bar\omega\tau}} \left[ C_+ (1- e^{-i p \tau}) - C_- (1- e^{i p \tau}) \right]/p.
\end{align}

From \eqref{g2def} we also find the expression for the second order field correlation function, which reads
\begin{align}
g^{(2)} (t) =& \frac12\bigg| 1  + \frac{4 \gamma^2 \big(1-\cos(\bar\omega\tau)\big)}{(\lambda^* +  i \gamma e^{-i\bar\omega\tau})^2}  \left[ I_1 (t)-\cos(\bar\omega\tau) I_0 (t)  \right] \bigg|^2,
\label{g2aa}
\end{align}
where we defined $I_1 (t ) = \big({I_0 (t-\tau) + I_0 (t+\tau)}\big)/2$ and
\begin{align}
I_0 (t) =& \frac{1}{2 (p \cos(p\tau) - i \lambda \sin(p\tau))} 
 \Big( e^{-i p (|t| +\tau)}  m^{-1} (p) -  e^{i p (|t| +\tau)} m^{-1} (-p)  
\nonumber \\
&+  \sum_{n=0}^{\infty} \Theta (|t|- n\tau ) [g_n^{(-)} (t) 
- g_n^{(+)} (t)] \Big),
\label{i0fin}   
\end{align}
with
\begin{align}
g_n^{(\pm)} (t) =&  ( i \gamma e^{i\bar\omega \tau})^n e^{\mp i p \tau} 
\frac{( \pm p +\lambda 
- i \gamma e^{i \bar\omega\tau} e^{\pm i p \tau})^2}{(\pm p +\lambda)^{n+1}}
 \left[ e^{\mp i p  (|t| - n\tau)} -e^{i \lambda (|t|- n\tau) } 
f_n^{(\pm)} (t) \right] , \\
f_n^{(\pm)} (t) =& \sum_{l=0}^n 
\frac1{l!}{[- i (|t |-n \tau) (\pm p +\lambda )]^{l}}.
\label{i1fin}  
\end{align}
Here the non-Markovian behavior manifests itself in the occurrence of additional contributions to $g^{(2)} (t)$ at every multiple of the delay time $\tau$, which create discontinuities in the derivative of $g^{(2)} (t) $. 

\subsection{V-level atom}

We now consider the atom as a V-level system, as represented in figure \ref{fig1}(b). In the single excitation subspace, the self-energy expresses as
\begin{align}
\Sigma (E) =&  - \frac{i}{2} \gamma (| e_1 \rangle \langle e_1 |+  | e_2 \rangle \langle  e_2 | ) + i \gamma e^{i (\bar\omega+E) \tau} | e_2 \rangle \langle  e_1|   .
\end{align}
From \eqref{eq:dyson}, it provides the dressed Green's function
\be
M (E) = \frac{1}{E+\lambda_{1}}| e_1 \rangle \langle e_1 |+\frac{1}{E+ \lambda_2}| e_2 \rangle \langle e_2|
  + \frac{i \gamma e^{i (\bar\omega+E)\tau} }{(E+\lambda_1)(E+ \lambda_2)}| e_2 \rangle \langle e_1 |  ,
\ee
where we defined $\lambda_{1,2} = \delta_{1,2} +i \gamma/2$. In this case we obtain from \eqref{eq:stnudef} the single-photon scattering phase $s =  \lambda_1^* \lambda_2^* /({\lambda_1 \lambda_2}).$
From \eqref{eq:defOmega}, we also get
\be
\bra{\nu'}\Omega(E)\ket{\nu} = \frac{\gamma}{2 \pi} \frac{\big(e^{i (\bar \omega + \nu)\tau/2}\ket{e_2} - e^{-i (\bar\omega+\nu)\tau/2}\ket{e_1} \big)  \big(e^{-i (\bar \omega + \nu')\tau/2}\bra{e_2} - e^{i (\bar \omega+\nu')\tau/2}\bra{e_1} \big) }{E-\nu' - \nu + i \eta}. \ee

We recall that each photon interacts only once with each transition in the scattering event, in contrast to the case of the two-level system where it can be reemitted towards the mirror an arbitrary number of times. As a consequence, the iterative equation  \eqref{eq:recWE} identically terminates after the third iteration, providing the expression
\begin{align}
 W^{\mathcal{P}}_{-\nu,0} (0) 
 \nonumber  
 &= \frac{\gamma}{2 \pi} \frac{\big(e^{i \bar \omega \tau/2}\ket{e_2} - e^{-i \bar\omega \tau/2}\ket{e_1} \big)  \big(e^{-i (\bar \omega - \nu)\tau/2}\bra{e_2} - e^{i (\bar \omega-\nu)\tau/2}\bra{e_1} \big) }{\nu }\nonumber \\
 &-  \Big(\frac{{\gamma}}{2\pi}\Big)^2 e^{i  \bar\omega \tau}  |e_2 \rangle \langle e_1 |  ( e^{-i\nu \tau /2} F_{1}^{(1)} (\nu)+e^{i\nu \tau /2} F_{1}^{(2)} (\nu)) \nonumber \\
 &-  \Big(\frac{\gamma}{2\pi}\Big)^3 e^{ i  \bar \omega\tau}  |e_2 \rangle \langle e_1 |     e^{i   \nu \tau /2} F_2 (\nu),
 \label{wcc0}
\end{align}
where
\begin{align}
F_1^{(\chi)} (\nu ) =&  -\frac{2 \pi i}{\lambda_\chi} \left[ \frac{e^{i \nu \tau} - e^{i \lambda_{\chi} \tau}}{\nu - \lambda_{\chi}} - \frac{e^{i \omega \tau} - 1}{\nu} \right],  \quad (\chi =1,2)  \label{F1w} \\
F_2 (\nu) =& \frac{4 \pi^2}{\lambda_1 (\nu - \lambda_2)} \left[ \frac{e^{i (\nu+ \lambda_1) \tau}-1}{\nu+ \lambda_1} - \frac{e^{i (\lambda_1 + \lambda_2) \tau}-1}{\lambda_1 + \lambda_2} - \frac{e^{i \nu \tau}-1}{\nu} + \frac{e^{i \lambda_2 \tau}-1}{\lambda_2} \right]. \label{F2w} 
\end{align}
Combining \eqref{eq:stnudef} and \eqref{wcc0} we deduce the inelastic part of the scattering as
\begin{align}
\hat{t} (\nu) =& - \frac{2 i\gamma^2}{\lambda_1 }  \frac{\lambda_2^{* 2}-\lambda_1^2}{\lambda_2^2-\lambda_1^2}   \frac{1}{\nu^2 - \lambda_{1}^2}  -  \left[   \frac{\lambda_1^{*2}}{\lambda_1^2} \frac{\gamma}{2\lambda_2}   -  \frac{ i \gamma^2 }{\lambda_1^2}  \frac{\lambda_2}{\lambda_1 + \lambda_2} \right. \nonumber \\
 & \left. +  \frac{i \gamma^2 }{\lambda_1^2}  \frac{\delta_2 \lambda_2}{\lambda_2^2 - \lambda_1^2}   +  \frac{ i \gamma^2}{2\lambda_1 \lambda_2 }  e^{i \lambda_2 \tau}  \left( \frac{\lambda_1^*}{\lambda_1} + \frac{ i \gamma \lambda_2 e^{i \lambda_1 \tau}}{ \lambda_1 (\lambda_1 + \lambda_2)} \right) \right] \frac{4i \gamma}{\nu^2 - \lambda_{2}^2} \nonumber \\
 &  -   \frac{2  i \gamma^2 }{\lambda_1 \lambda_2 } \left( \frac{\lambda_1^*}{\lambda_1} + \frac{ i \gamma \lambda_2 e^{i \lambda_1 \tau}}{ \lambda_1 (\lambda_1 + \lambda_2)} \right) \left[ \frac{\lambda_2^* - \lambda_1}{\lambda_2 - \lambda_1}  \frac{\lambda_1 \cos \nu \tau - i \nu \sin \nu \tau}{\nu^2 - \lambda_1^2} \right. \nonumber \\
 & \left. +  \frac{i \gamma}{2\lambda_2 }  
  \frac{\lambda_1 +   \lambda_2 }{\lambda_2 - \lambda_1 }  \frac{\lambda_2 \cos \nu \tau -i \nu \sin \nu \tau}{\nu^2 - \lambda_2^2}   + \frac{\delta_2}{\lambda_2} \frac{\lambda_2 \cos \nu \tau + i \nu \sin \nu \tau}{\nu^2 - \lambda_2^2} \right] .
  \label{tccc}
\end{align}
Note that in this expression there is no dependence on the phase factor $e^{i\bar\omega\tau}$, since it can be absorbed in the re-definition of the atomic excited states, but there is a nontrivial dependence on $\tau$. This means that the distance to the mirror becomes a relevant parameter only in the non-Markovian regime.

With the expression of $\hat{t}(\nu)$ we find readily the elastic and inelastic power spectra in \eqref{Sel_inel}. The second order field correlation is finally evaluated from \eqref{g2def} in terms of 

\begin{align}
 \frac{1}{2 \pi} &\int d \nu\,  \hat{t} (\nu) \cos(\nu t) \nonumber 
= \frac{\gamma^2 }{\lambda_1^2}  \frac{\lambda_2^{* 2}-\lambda_1^2}{\lambda_2^2-\lambda_1^2} e^{i \lambda_1 t}  \\ &+ \frac{2 \gamma }{\lambda_{2}} \left[   \frac{\lambda_1^{*2}}{\lambda_1^2} \frac{\gamma}{2 \lambda_2}   -  \frac{ i \gamma^2 }{\lambda_1^2}  \frac{\lambda_2}{\lambda_1 + \lambda_2} \right.  \nonumber  
+  \frac{ i \gamma^2 }{\lambda_1^2}  \frac{\delta_2 \lambda_2}{\lambda_2^2 - \lambda_1^2} 
  \left.+  \frac{ i \gamma^2}{2\lambda_1 \lambda_2 }  e^{i \lambda_2 \tau}  \left( \frac{\lambda_1^*}{\lambda_1} + \frac{ i \gamma \lambda_2 e^{i \lambda_1 \tau}}{ \lambda_1 (\lambda_1 + \lambda_2)} \right) \right] e^{i \lambda_2 t} \nonumber \\
&  +   \frac{\gamma^2 }{\lambda_1 \lambda_2 } \left( \frac{\lambda_1^*}{\lambda_1} + \frac{ i \gamma \lambda_2 e^{i \lambda_1 \tau}}{ \lambda_1 (\lambda_1 + \lambda_2)} \right) \left[ \frac{\lambda_2^* - \lambda_1}{\lambda_2 - \lambda_1}  \Theta (t - \tau) e^{i \lambda_1 (t - \tau)} \right. \nonumber \\
&  \left. +  \frac{i \gamma }{2\lambda_2 }  
  \frac{\lambda_1 +   \lambda_2 }{\lambda_2 - \lambda_1 }  \Theta (t - \tau) e^{i \lambda_2 (t - \tau)}  + \frac{\delta_2}{\lambda_2}  e^{i \lambda_2 (\tau+ t)} +  \frac{\delta_2}{\lambda_2}  \Theta (\tau- t) e^{i \lambda_2 (\tau- t)} \right].
  \label{gcc_ic}
\end{align}
Once again, this function also features a discontinuity of the first derivative at $t=\tau$, which shows the effect of the delay. However, there is no discontinuity for $t>\tau$ as here the photon immediately leaves the feedback loop after this delay.

\section*{References}

\bibliographystyle{iopart-num.bst}

\providecommand{\newblock}{}

\end{document}